\documentclass[%
 aip,
 amsmath,amssymb,
 reprint,%
]{revtex4-1}

\usepackage{graphicx}
\usepackage{dcolumn}
\usepackage{bm}

\usepackage[utf8]{inputenc}
\usepackage[T1]{fontenc}
\usepackage{mathptmx}
\usepackage{etoolbox}

\makeatletter
\def\@email#1#2{%
 \endgroup
 \patchcmd{\titleblock@produce}
  {\frontmatter@RRAPformat}
  {\frontmatter@RRAPformat{\produce@RRAP{*#1\href{mailto:#2}{#2}}}\frontmatter@RRAPformat}
  {}{}
}%

\usepackage{pdfpages}

\makeatletter
\AtBeginDocument{\let\LS@rot\@undefined}
\makeatother

\makeatother
\begin{document}

\preprint{AIP/123-QED}

\title[Thermal conductivity in modified oxide glasses is governed by modal phase changes]{Thermal conductivity in modified oxide glasses is governed by modal phase changes}
\author{Philip Rasmussen}
\author{Søren S. Sørensen$^{*,}$}%
 \email{soe@bio.aau.dk}
\affiliation{ 
Department of Chemistry and Bioscience, Aalborg University, Fredrik Bajers Vej 7H, 9220 Aalborg, Denmark
}%

\date{\today}

\begin{abstract}
The thermal conductivity of glasses is well-known to be significantly harder to theoretically describe compared to crystalline materials. Because of this fact, the fundamental understanding of thermal conductivity in glasses remain extremely poor when moving beyond the case of simple glasses, e.g., glassy SiO$_2$, and into so-called 'modified' oxide glasses, that is, glasses where other oxides (e.g. alkali oxides) have been added to  break up the network and alter e.g. elastic and thermal properties. This lack of knowledge is apparent despite how modified glasses comprise the far majority of known glasses. In the present work we study an archetypical series of sodium silicate ($x\text{Na}_2\text{O}\text{-}(100\text{-}x)\text{SiO}_2$) glasses. Analyses of modal contributions reveal how  increasing Na$_2$O content induces increasing vibrational localization with a change of vibrations to be less ordered, and a related general decrease in modal contributions to thermal conductivity. We find the vibrational phases (acoustic vs. optical) of sodium vibrations to be relatively disordered compared to the network-forming silicon and oxygen species, explaining how increasing Na$_2$O content decreases thermal conductivity.  Our work sheds new light on the fundamentals of glassy heat transfer as well as the interplay between thermal conduction and modal characteristics in glasses.
\end{abstract}

\maketitle

\section{Introduction}

Thermal conductivity ($\kappa$) is a fundamental materials property with relation to both specific applications in e.g. insulating materials and thermoelectrics, as well as fundamental linkages to the underlying vibrational character of materials~\cite{Peierls1929,Kittel1949,Zeller1971,Allen1989}. However, for the property of thermal conductivity, big differences between solid materials families exist. The perhaps biggest difference being that between crystalline and amorphous/glassy materials, where crystals generally feature larger $\kappa$ than their amorphous counterparts~\cite{Kittel1949,Zeller1971,Soerensen2020}. While glasses (i.e., amorphous materials exhibiting a glass transition~\cite{Zanotto2017}) is a somewhat diverse materials family, the most industrially relevant glasses are the oxides with practical applications in bulk applications such as construction and screens, yet they also see use in more specialized applications as e.g. electrical insulation materials in microchip manufacturing~\cite{zhou2020thermal,Varshneya2013} due to their electrically-insulating properties. In detail, oxide glasses are chemically distinct by being composed of network formers (e.g., SiO$_2$, GeO$_2$, B$_2$O$_3$, P$_2$O$_5$)~\cite{zachariasen1932atomic}, often in combination with so-called network modifiers (alkali, alkaline earth, and transition metal oxides)~\cite{Varshneya2013,greaves1985exafs}. Structurally, the network formers comprise the back-bone (e.g. Si-O-Si) of the glass, while the introduction of the network modifier breaks up the backbone structure by introducing so-called non-bridging oxygens (e.g., Si-O$^-$~Na$^+$) where the metal-ion charge balances the created negatively charged non-bridging oxygens~\cite{Varshneya2013,greaves1985exafs}. 

Generally, the range of thermal conductivity in both organic and inorganic glass materials is rather narrow (typically $~$0.1-2.0~W~m$^{-1}$~K$^{-1}$) compared to that of crystals (0.1-$10^4$~W~m$^{-1}$~K$^{-1}$). For oxides specifically, both pure network formers as well as modified oxide glasses are experimentally rather well-studied, typically covering a $\kappa$-range of 0.4-1.4~W~m$^{-1}$~K$^{-1}$~\cite{Ratcliffe1963}, with pure network formers being in the range of 0.5-1.4~W~m$^{-1}$~K$^{-1}$~\cite{Ratcliffe1963}. However, while empirical rules for tweaking $\kappa$ in oxide glasses are quite vast~\cite{Ratcliffe1963,Vavilov1982,pye2005properties}, the underlying fundamental understanding of the heat propagation mechanism is much less studied, with the majority of studies focusing on pure SiO$_2$~\cite{Jund1999,Martin2022} and, to our knowledge, only very few studies on modified oxides exist~\cite{Soerensen2019,SORENSEN2022104160}. This lack of fundamental understanding is largely due to how the the lack of long-range structural order complicates the theoretical description of heat propagation which is otherwise applicable for crystals (and has been for decades)~\cite{McGaughey2006} as well as the lack of reliable interatomic potentials capable of describing the vibrational characteristicss of these systems. However, very recently, two theories of estimating thermal conductivity were unified to comprehend both amorphous and crystalline materials as well as structures in between these limits~\cite{Peierls1929,Allen1989,Simoncelli2019,Isaeva2019} based on lattice dynamics to the third order. This has allowed a significant broadening in the possible studies capable of being performed. Yet despite the improved theoretical description, studies are still limited and a general description beyond that of the simplest single-oxide case is lacking.

To overcome this, in the present work we showcase how an existing classical force field is capable of  reproducing the structural and vibrational features of several archetypical sodium silicate glasses (in the series of $x\text{Na}_2\text{O}\text{-}(100\text{-}x)\text{SiO}_2$ with $x$~=[20-35]). Using the recent advances in modelling thermal conductivity, we furthermore show how the used force field provides a superior estimation of $\kappa$ in the given systems in very good agreement with experimental findings. We supplement this estimation with analyses of mode localization and mode character to provide general knowledge on how addition of modifying oxides to pure network forming glasses affect the fundamental modal types and how these contribute to the thermal conductivity in modified oxide glass systems. Ultimately, our results provide new knowledge on the fundamentals of the connection between heat transfer and structure in archetypical modified glass systems and may aid in future development of glasses of ultra-high or -low thermal conductivity.

\section{Methods}

\subsection{Glass preparation}
The archetypal sodium oxide glass system $x\text{Na}_2\text{O}\text{-}(100\text{-}x)\text{SiO}_2$ with $x$~=~\{0, 10, 15, 20, 25, 30, 33.3, 35, 40\} has been investigated by employing the well-established potential of Teter with the potential form,

\begin{equation}
    U(r_{ij}) = \frac{q_iq_j}{4\pi r_{ij} \epsilon_0}+A_{ij} \text{exp}\left(-\frac{r_{ij}}{\rho_{ij}}\right)-\frac{C_{ij}}{r^6_{ij}}+\frac{D_{ij}}{r_{ij}^{24}},
\end{equation}

\noindent

where $U$ is the interatomic potential energy, $q$ is atomic charge, $r$ is the interatomic separation, $\epsilon_0$ is the permittivity of vacuum, while $i$ and $j$ denote atomic species.
Here, the first term is the Coulombic term which accounts for the charge-charge interactions between atoms. The second and third term are the short range Buckingham potential with parameters $A_{ij}$, $\rho_{ij}$ and $C_{ij}$ used as fitted by Teter~\cite{Cormack2002,Du2015}. The last term is a short-repulsive term added to prevent the "Buckingham catastrophe" at very short distances and high temperatures~\cite{Carr2016}. The correction is illustrated in Figure~S1 in the Supplementary Material showing the correction provided to remove the infinitely deep potential well at short separating and hence prevent unphysical forces and ultimately crashing simulations. Potential parameters used in simulations are provided in Table~\ref{teter_pot_parameters}. The short-range interactions were evaluated with a cutoff of 8~Å while the long-range Coulombic interactions were evaluated in real space below 12~Å and using the PPPM algorithm using an accuracy of $10^{-5}$ above 12~Å. Glasses were prepared using the melt-quench procedure using the Large-scale Atomic/Molecular Massively Parallel Simulator (LAMMPS) package~\cite{Plimpton1995} with a fixed timestep of 1.0~fs

Box sizes of 3000 atoms were employed in simulations to study structural features, elastic properties and various vibrational characteristics to limit errors from finite size effects. The number of Si, O and Na atoms in each glass were fixed according the corresponding molar composition of the glasses investigated, and box sizes were estimated according to experimental glass densities (see Figure~S2 in the Supplementary Material) adapted from various studies~\cite{Du2004,White1977,young}. Detailed simulation cell information is provided in Table~S1 in the Supplementary Material. 

\begin{table}[!htbp]
\caption{Interatomic potential parameters used in simulation of sodium silicate glasses.}
\centering
{\begin{tabular}{ccccc}
\hline
 Interaction & $A$ (eV) & $\rho$ (\AA) & $C$ (eV~\AA$^6$)& $D$ (eV~\AA$^{24}$)\\
\hline
 Na$^{+0.6}$-O$^{-1.2}$ & 4383.7555 & 0.243838 & 30.70 & 1\\
O$^{-1.2}$-O$^{-1.2}$ & 2029.2204 & 0.343645 & 192.58 & 113\\
 Si$^{+2.4}$-O$^{-1.2}$ & 13702.905 & 0.193817 & 54.681 & 29\\
\hline
\end{tabular}}
\label{teter_pot_parameters}
\end{table}

Initially, 3000 atoms were placed randomly in a simulation box separated by a minimum of 1.6~Å to avoid unphysical dynamics at the initiating MD steps, followed by an energy minimization with an energy and force tolerance of $10^{-10}$. Glasses were then equilibrated at 5000~K for 2~ns to ensure complete memory loss of the initial atomic configuration. A subsequent linear cooling was performed to 300~K with a cooling rate of 10~K~ps$^{-1}$. Finally, glasses were relaxed for 100~ps where atomic trajectories were averaged and analyzed over 100~ps. All simulations were performed in the NVT ensemble.

Short range structural features such as coordination numbers, angular distribution functions and bond lengths were estimated from the final part of the quenching procedure employing the partial radial distribution functions computed with the formula,

\begin{equation}
     g_{i j}(r) = \frac{\langle n_{i j}(r)\rangle dr}{4\pi r^2dr \rho_0 C_{j}},
\end{equation}\label{rdf}

\noindent
where $\rho_0$ is the average atomic number density and $n_{ij}$ is the average number of particles between $r+dr$. The angular distribution and coordination numbers were determined by investigating the first coordination shells of different atomic species where a cutoff of 2.2~Å and $\sim$3.3~Å were used for the Si-O and Na-O interactions, respectively.

\subsection{Structural characterization}

In order to investigate the ability of the potential of Teter to predict realistic medium-range structure in sodium silicate glasses, we computed the total structure factor to compare it to experimental scattering data. To do so, first, the partial structure factors were computed,

\begin{equation}
     S_{ij}(Q) = 1+\rho_0 \int_0^R 4\pi r^2(g_{ij}-1) \frac{\sin(Qr)}{Qr} \frac{\sin(\pi r/R)}{\pi r/R}dr,
\end{equation}

\noindent
where $Q$ is the wave vector, $\rho_0$ is the average atomic density, $R$ is the real space maximum value of integration, $g_{ij}$ is the partial radial distribution function and ${\text{sin}(\pi r/R)/(\pi r/R)}$ is a Lorch-type function used to reduce effects from having finite box sizes. The neutron total structure factor was computed from the partial structure factors,

\begin{equation}
      S_{N}(Q) = \left( \sum^n_{i=1}c_ib_i\right)^{-2} \sum^n_{i,j=1}c_ic_jb_ib_jS_{ij}(Q),
\end{equation}
\noindent
where $c_i$ is the fraction of atom species $i$ and $b_i$ is neutron scattering length of atomic species $i$. The neutron scattering lengths are $3.63\cdot10^{-15}$~m, $5.803\cdot10^{-15}$~m and $4.1491\cdot10^{-15}$~m for sodium, oxygen, and silicon, respectively~\cite{sears1992neutron}. Simulated structure factors were evaluated by comparing with experimental structure factors from experimental neutron scattering experiments by employing the $R_{\chi}$-factor as proposed by Wright~\cite{Wright1993},

\begin{equation}
       R_\chi = 100\%~\sqrt{\frac{\sum_r[S_N(Q)^{\text{exp}}-S_N(Q)^{\text{sim}}]^2}{\sum_r[S_N(Q)^{\text{exp}}]^2}},
\end{equation}

\noindent
providing a quantitative measure of the agreement between simulated and experimental structure. Generally, a simulated structure with $R_\chi$<10\% is considered to have a good agreement with experiment.

\subsection{Mechanical properties}

To further validate the ability of the Teter potential~\cite{Cormack2002,Du2015} to predict structure and properties, the elastic properties of simulated sodium silicate glasses were estimated by displacing the simulation boxes with a finite difference in 6 perturbations, namely in in the $xx$, $yy$, $zz$, $xy$, $xz$ and $yz$ directions, while measuring the 6 components of the pressure tensor, i.e. the directional pressure components. The simulation boxes were displaced in steps of 0.1~\% of the current box length for the tensile strain and in steps of 0.1~Å for the shear strain. Performing linear regression in the elastic region yielded the elastic constants $C_{11}$ and $C_{44}$, which are used to compute the elastic constant $C_{12}$, bulk ($B$), Young's ($E$) and shear ($G$) moduli and Poisson's ratio ($v$),

\begin{equation}
    C_{12}=C_{11}-2C_{44},
\end{equation}

\begin{equation}
    B=\frac{C_{11}+2C_{12}}{3},
\end{equation}

\begin{equation}
    E= \frac{(C_{11}-C_{12})(C_{11}+2C_{12})}{C_{11}+C_{12}},
\end{equation}
    
\begin{equation}
    G=\frac{C_{11}-C_{12}}{2}=C_{44},
\end{equation}

\begin{equation}
     v=\frac{C_{12}}{C_{11}+C_{12}}.
\end{equation}

\subsection{Thermal conductivity}

To investigate the ability of the Teter potential to predict vibrational and thermal properties in sodium silicate glasses, we compute the thermal conductivity of simulated glasses by employing the recently developed Quasi Harmonic Green-Kubo (QHGK) method~\cite{Isaeva2019,Simoncelli2019} as implemented in the $\kappa$ALD$o$ software~\cite{Barbalinardo2020}. The method is a unification of the phonon gas model~\cite{Peierls1929} and the theory of heat transport in glasses provided by Allen and Feldman~\cite{Allen1989}, allowing for a generalized description of both crystalline and amorphous materials~\cite{Isaeva2019}, where $\kappa$ is given as,

\begin{equation}\label{QHGK}
     \kappa_{\alpha\beta} = \frac{1}{V}\sum_{nm} C_{nm} v^\alpha_{nm} v^\beta_{nm} \tau^\circ_{nm},
\end{equation}


\noindent
where $V$ is the simulation cell size, $n$ and $m$ are mode indexes, $\alpha$ and $\beta$ are cartesian components, $C_{nm}$ is the generalized heat capacity, $v^\alpha_{nm}$ is the generalized phonon group velocities and $\tau^\circ_{nm}$ is the generalized phonon relaxation times. Then, Eq.~\ref{QHGK} conveniently provides the per mode thermal conductivity of simulated glasses. In order to estimate thermal conductivity with the QHGK method, the second and third-order IFC's were computed with the LAMMPS package (using the built-in finite displacement method using displacements of $10^{-6}$~Å), and subsequently imported into $\kappa$ALD$o$ along with a melt-quenched energy minimized glass structure file. Computation of the third-order IFC's takes a high computational effort since the number of matrix elements scales like (3$N$)$^3$ where $N$ is the number of atoms. Hence, box sizes of 600 atoms were employed for the QHGK calculations to significantly reduce the computational expenses. Simulation cell information of 600 atom systems can be found in Table~S2 in the Supplementary Material. Moreover, evaluating the long-range Coulombic forces in reciprocal space with the PPPM algorithm was found to yield unphysically low thermal conductivity values and a significant increase in computational cost. Instead, similarly to a previous study~\cite{SORENSEN2022104160}, the Coulombic damped shift force model~(DSF) was employed for simulations of the 600 atom glasses and computation of the second and third-order IFC's, using a damping factor of 0.3 and a cutoff of 11~Å, which conveniently reduces the third-order IFC file sizes and computational demand. Thermal conductivity estimations were then performed at 300~K, while also taking quantum effects (i.e., the Debye scaling of the heat capacity) into account.

\subsection{Vibrational Characteristics}

Further investigation of vibrational and thermal properties was conducted by evaluating the second order interatomic force constants (IFC). The second order IFC's ($\Phi_{i\alpha j\beta}^{''}$) were then used to obtain the dynamical matrix ($\mathbf{D})$ which is the second IFCs rescaled by the atomic masses, that is,

\begin{equation}\label{dynamical_matrix}
    \mathbf{D} = \frac{\Phi_{i\alpha j\beta}^{''}}{\sqrt{m_i m_j}}, 
\end{equation}

\noindent
where $m_i$ is the atomic mass of atom $i$. Then, from the dynamical matrix, the mode eigenfrequencies ($\Omega$) and eigenvectors ($\mathbf{e}$) were obtained by solving the following eigenvalue equation,

\begin{equation}\label{eq.eigenvectors}
    \mathbf{e}\cdot \Omega= \mathbf{D} \cdot \mathbf{e},
\end{equation}

where $\mathbf{e}$ contains $3N$ column vectors ($\mathbf{e}_i$), where $N$ is the number of atoms, which provides information about the relative motion of all atoms in each eigenmode, and $\Omega$ is a square matrix with diagonal elements of the squared angular modal eigenfrequencies,

\begin{equation}
       \Omega =
     \renewcommand{\arraystretch}{0.8}
       \begin{bmatrix}
\omega^2_1&  &  &  &  &  \\
 & \omega^2_2 &  & & & \\
 &  & \omega^2_3& & & \\
 &  &  & . & & \\
 &  &  & & . & \\
 &  &  & & & \omega^2_{3N}  \\
     \end{bmatrix}
\end{equation}

\noindent
which are readily converted into ordinary frequencies by $\omega^2/2\pi$. We then bin the $3N$ eigenfrequencies into a histogram to obtain the vibrational density of states ($g(\omega$)),

\begin{equation}
    g(\omega) = \sum^{3N}_n (\omega_n - \omega)\delta,
\end{equation}

\noindent
where $\delta$ is the Dirac delta. To further investigate the modal vibrational characteristics, we exploit the modal eigenvector normalization,

\begin{equation}\label{eigenvector_normalization}
\begin{split}
\sum^N_i \vec{\mathbf{e}_{i}}(n)\cdot \vec{\mathbf{e}_{i}}(n) =  \sum^{N_{Na}}_i \vec{\mathbf{e}_{i}}&(n)\cdot \vec{\mathbf{e}_{i}}(n) +\\ \sum^{N_{Si}}_i \vec{\mathbf{e}_{i}}(n)\cdot \vec{\mathbf{e}_{i}}(n)& + \sum^{N_{O}}_i \vec{\mathbf{e}_{i}}(n)\cdot \vec{\mathbf{e}_{i}}(n) =  1,
\end{split}
\end{equation}

\noindent
where $i$ denotes the atomic index number, $n$ is the modal index number and $N_{\text{Na/O/Si}}$ is the number of sodium, oxygen, and silicon atoms, respectively. The inner product of each eigenvector essentially yield weighting factors for each atom and atomic species, providing a measure of the vibrational participation of atomic species in the collective modal motion. This is utilized to obtain the partial $g(\omega)$~\cite{bauchy}

\begin{equation}
    g_\alpha(\omega) = g(\omega) \sum_{i\in\alpha} |\vec{\mathbf{e}_{i}}(n)|^2,
\end{equation}

\noindent

where $\alpha$ denotes atomic species with $\alpha$ $=$ (Si, Na, O, BO, NBO, ${\text{Si}^{\text{Q}^1}}$, ${\text{Si}^{\text{Q}^2}}$, ${\text{Si}^{\text{Q}^3}}$, ${\text{Si}^{\text{Q}^4}}$), where BO and NBO is bridging and non-bridging
oxygen, respectively, and ${\text{Si}^{\text{Q}^n}}$ is silicon atoms in ${\text{Q}^n}$ species of different degrees of polymerization ($n$ related to the number of bridging-oxygens attached to the Si atom).

We further utilize the eigenvector normalization from Eq.~\ref{eigenvector_normalization} to compute the per mode participation ratio ($PR$) to estimate the per type participation in each mode~\cite{elliott1990physics},

\begin{equation}
    PR (n) = \left(N \sum_i^N |\vec{\mathbf{e}_{i}}(n)|^4\right)^{-1},
\end{equation}

\noindent
where all individual atomic participation ratio values for atoms $i$ are summed over the entire $n$-th mode. Practically, the limit of $PR=1$ shows an equal contribution from all atoms to the vibrations while the limit of $1/N$ indicates a mode of only 1 atom contributing to the modal vibration.

To gain further insight into the participation ratio in simulated glasses we employ the atomic participation ratio ($APR$)~\cite{Pailhs2014},

\begin{equation}
    APR_\alpha (n) = \frac{|\vec{\mathbf{e}_{i}}(n)|^2}{\left(N \sum_{i\in\alpha}^{N_i} |\vec{\mathbf{e}_{i}}(n)|^4\right)^{\frac{1}{2}}},
\end{equation}
\noindent
which is an atomic decomposition of the participation ratio, providing the participation ratio of individual atomic species ($APR_{\alpha}$) that satisfies $PR (n)= \sum_{i\in\alpha}$ ($APR_{\alpha}(n))^2$.

Next, to distinguish localized modes from delocalized modes we compute the inverse participation ratio, simply by computing the inverse of the PR, 

\begin{equation}
    IPR(n) = (PR(n))^{-1}, 
\end{equation}
\noindent
which provides a measure of the frequency range at which the mobility edge appears and localized modes are prevalent.

To further characterize the modal motion, the phase quotient ($PQ$), originally proposed by Bell and Hibbins-Butler\cite{Bell1975}, was used to distinguish between acoustic ($PQ=1$) and optical ($PQ=-1$) modes, which provides a measure of whether nearest neighbor atoms vibrate mostly in phase or out of phase in each mode. Specifically, $PQ$ is given as,

\begin{equation}
    PQ(n) = \frac{\sum_{m} \vec{\mathbf{e}_{i}}(n)\cdot \vec{\mathbf{e}_{j}}(n)}{\sum_{m}|\mathbf{e}_{i}(n)\cdot \vec{\mathbf{e}_{j}}(n)|}.
\end{equation}
\noindent

We note that the terms 'acoustic' and 'optical' are commonly used to define fully in-phase and out of phase modes from the dispersion relation of crystalline phonons, yet, the terminology still somewhat applies and conveniently provides a quantitative measure of the average degree of phase motion of modes in glasses. We further study the atomic motion and its relation to nearest neighbors by investigating the atomic phase quotient ($APQ$) of different atomic species,

\begin{equation}
    APQ(n) = \frac{\sum_{m} \vec{\mathbf{e}_{i,\alpha}}(n)\cdot \vec{\mathbf{e}_{j}}(n)}{\sum_{m}|\vec{\mathbf{e}_{i,\alpha}}(n)\cdot \vec{\mathbf{e}_{j}}(n)|},
\end{equation}

\noindent
which in nature is similar to the $PQ$ but describes the direction of eigenvectors of each nearest neighbor atom adjacent to specific atomic species.

\section{Results and discussion}

\subsection{Structural and mechanical characterization}

\begin{figure}[!htb]
    \centering
    \includegraphics[width = 8.6cm]{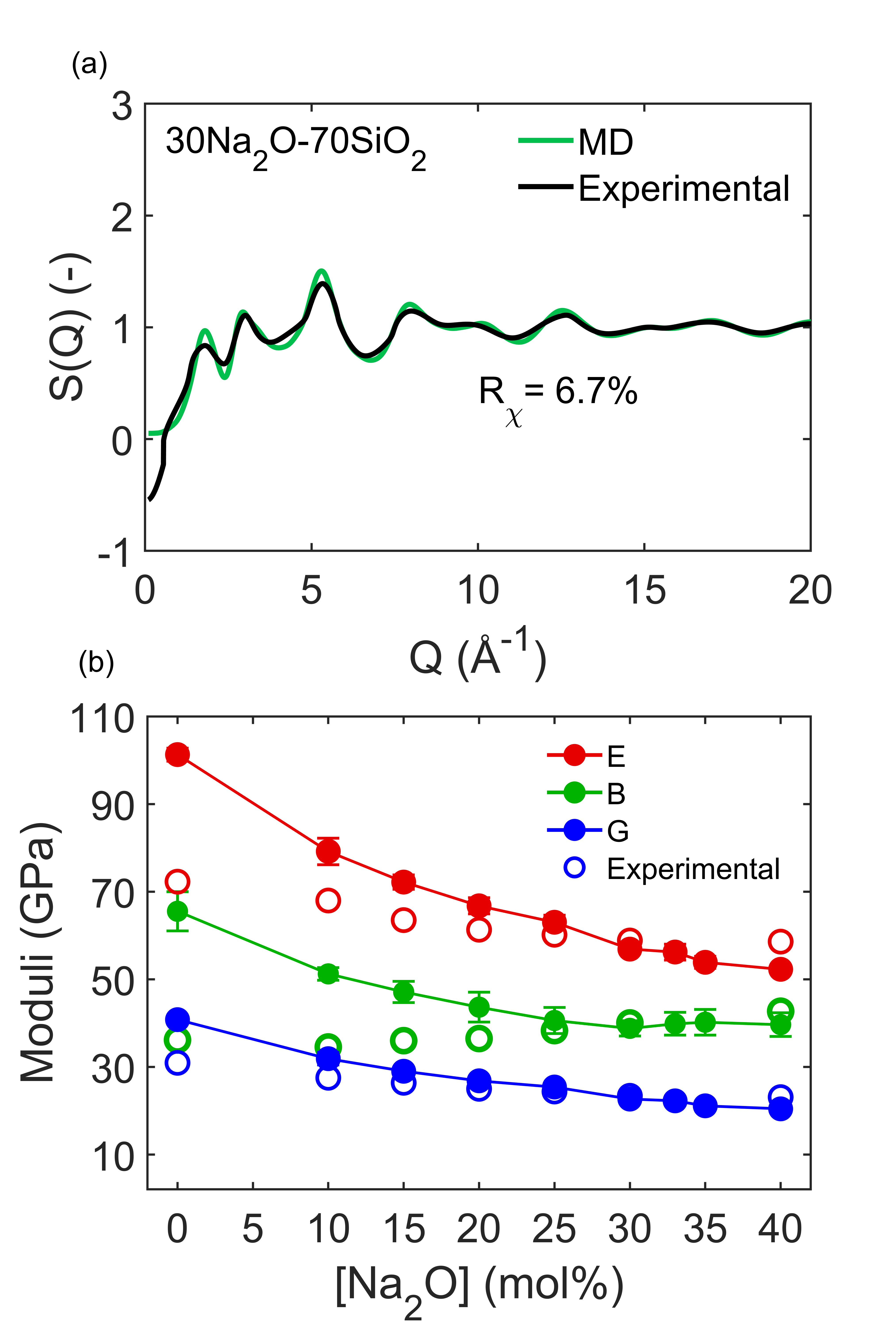}
    \caption{(a) Neutron-weighted structure factor of the studied glass with a composition of 30Na$_2$O-70SiO$_2$ from molecular dynamics (MD) simulations (green) as well as from experiments (black) showing a very good agreement and (b) Young's~(E), Bulk~(B) and Shear~(G) moduli of simulated glasses (filled circles) compared to experimental values (open circles). The experimental structure factor data is from Ref.~\cite{Fbin2007} and moduli data are from Ref.~\cite{Zhao2012}}
    \label{fig1_sq_moduli}
\end{figure}

In this work we employ the force field of Teter~\cite{Cormack2002,Du2015} which has previously been shown to provide a good description of the structure and mechanics in a number of alkali silicates~\cite{Du2005_1,Du2005_2,Du2006,Du2011,Du2015,Lodesani2020,Yu2017}. We first prepare $x\text{Na}_2\text{O}\text{-}(100\text{-}x)\text{SiO}_2$ glasses with $x$~=~\{\text{0, 10, 15, 20, 25, 30, 33.3, 35, 40}\} to make an initial assesment of properties in a very broad compositional range. First, to probe the structural agreement with experiments we present the calculated and measured neutron structure factor (experimental $S$($Q$) from Ref.~\cite{Fbin2007}) in Figure~\ref{fig1_sq_moduli}a for a 30Na$_2$O-70SiO$_2$ glass as well as a similar comparison for pure SiO$_2$ in Figure~S3 in the Supplementary Material. 

Notably, by comparing the experimental and simulated $S(Q)$ using the $R_{\chi}$ factor introduced by Wright~\cite{Wright1993}, both 30Na$_2$O-70SiO$_2$ and SiO$_2$ glasses see values well below 10\%, indicating how the simulations provide a very good structural depiction of the experimental glass in good agreement with previous reports of the present pair potential~\cite{Cormack2002,Du2015}. We supplement the structure factor comparisons with calculated $Q^n$-distributions (where $n$ indicates the fraction of bridging oxygens per Si tetrahedra), bond-angle distributions (Si-Bridging oxygen-Si,  O-Si-O as well as various Na-O related bond angles), total and partial structure factors as well as estimations of the fraction of bridging and non-bridging oxygen species for all studied compositions. These data are presented in Figures~S4-11 in the Supplementary Material. While not all parameters are accessible experimentally, for those available we find good agreement between the simulated glasses and experiments. For example, we identify O-Si-O bond centered around angles of $\sim$109$^{\circ}$ (Figure S6) as expected by a tetrahedral configuration, overall good qualitative agreement with $Q^n$ distributions (Figure S11a), and an expected decrease in bridging oxygen with increasing Na$_2$O content (Figure S11b).

Next, we probe the mechanical properties of the system by performing displacements in the tensile and shear directions to record the elastic constants of the system and ultimately to estimate the elastic moduli~\cite{Gersten2001} (see Methods for details). The results are shown in Figure~\ref{fig1_sq_moduli}b. From this analysis we find decreasing Young's ($E$), bulk ($B$) and shear ($G$) moduli with increasing Na$_2$O content. This is in good agreement with experiments for $E$ and $G$, but disagrees with the somewhat constant experimental $B$ (open green symbols in Figure~\ref{fig1_sq_moduli}b as obtained from Ref.~\cite{Zhao2012}) in the studied modifier concentration range. This disagreement is likely also the cause of the deviation between simulated and experimental Poisson's ratio although the general increase with increasing Na$_2$O trend is reproduced (see Figure~S12 in the Supplementary Material). In summary, we generally find good agreement between experimental and simulated mechanical properties, especially in the 20-35~mol\% Na$_2$O range, highlighting how the force field performs best in this concentration range. Given how the present study will focus on thermal transport, a property somewhat related to mechanical properties, the remaining part of this work will focus on  $x\text{Na}_2\text{O}\text{-}(100\text{-}x)\text{SiO}_2$ glasses with Na$_2$O contents of $x$~=~\{\text{20, 25, 30, 33.3, 35}\}.

\subsection{Vibrational features}\label{section:vibrational_features}

\begin{figure*}[!t]
    \centering
    \includegraphics[width = .9\textwidth]{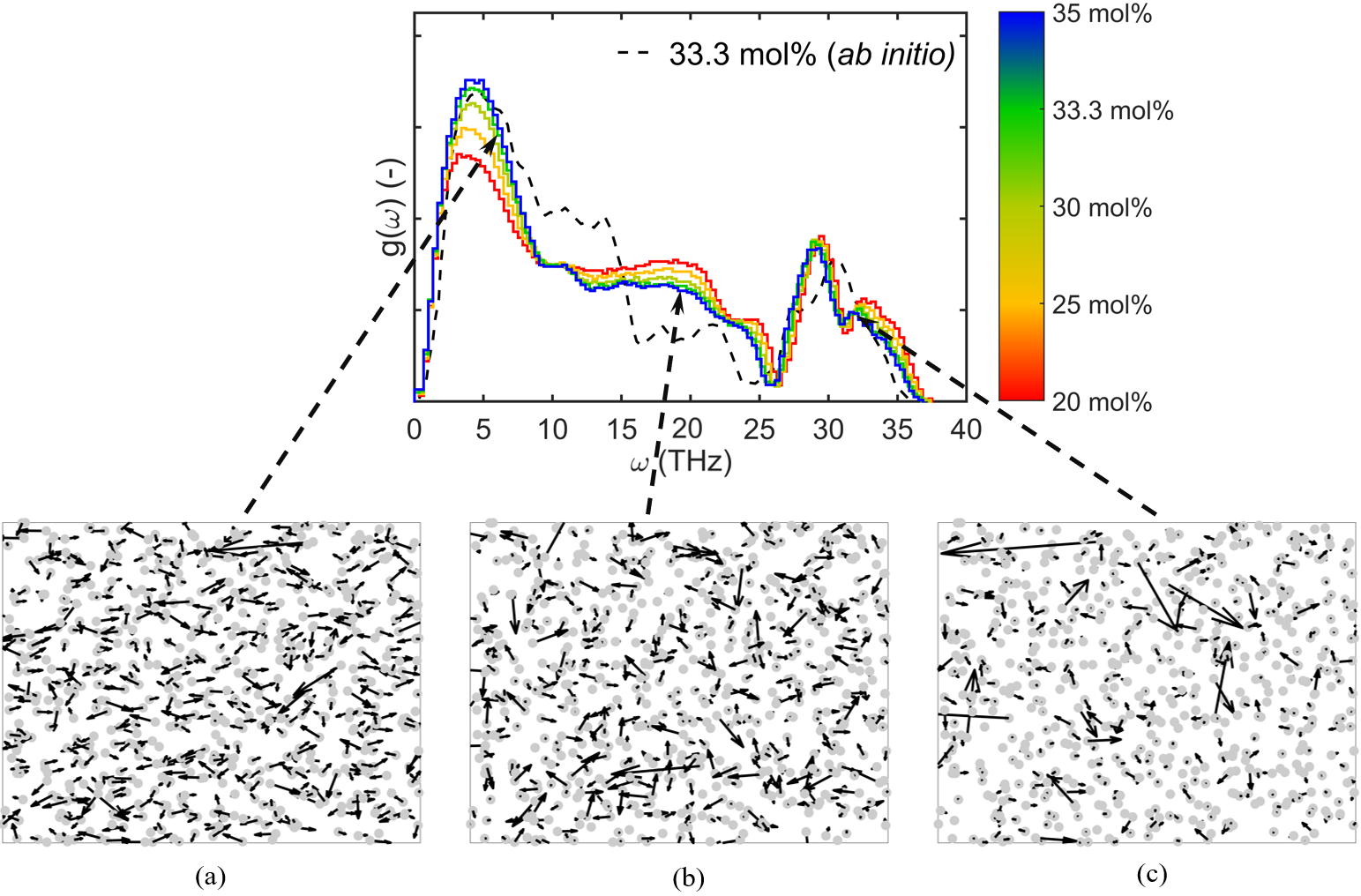}
    \caption{(Top) Vibrational density of states ($g(\omega)$) of simulated sodium silicate glasses with increasing sodium content from 20 to 35 mol\% Na$_2$O as well as the $g(\omega)$ from \textit{ab initio} simulations as performed in Ref.~\cite{Kilymis2019}, and (bottom) illustrations of eigenvectors of modes in three different frequency ranges of $g(\omega)$, i.e. (a) a low-frequency mode at $\sim$7~THz, (b) a mid-frequency mode at $\sim$20~THz and (c) a high-frequency mode at $\sim$32~THz, respectively.}
    \label{fig2_vdos}
\end{figure*}

While a number of properties have previously been tested using the Teter potential~\cite{Cormack2002,Du2015}, the force field has mainly been employed due to its superior modelling of structural and mechanical characteristics of modified alkali silicate glasses~\cite{Du2004,Du2005_1,Du2005_2,Du2006,liu2020searching,Mantisi2015}. A few studies have probed the total and partial vibrational density of states (VDOS)~\cite{Mantisi2015,bauchy} however with emphasis on the density/pressure response of glasses at or near the Na$_2$O-2SiO$_2$ composition. As such, to our knowledge the potential have never been used to test the thermal conductivity nor any detailed vibrational characteristics of glasses. In the present work we perform a detailed vibrational analysis of the compositions of interest as well as provide the first use of the potential to simulate the thermal conductivity of these glass types. First, we compute the total vibrational density of states ($g(\omega)$, see Methods) for a range of compositions as presented in Figure~\ref{fig2_vdos}.


Comparing the simulated data of the glasses produced using the Teter potential to \textit{ab initio} simulations (from Ref.~\cite{Kilymis2019}, See Figures~S13-14) we find good agreement. For more, our calculations agree well with previous estimations of the VDOS from the same potential~\cite{bauchy}. This generally suggests that our simulations are capable of decently describing the vibrational character of the studied glasses and thus supports the usage of the potential to predict other vibrationally-related properties, e.g. thermal conductivity. In addition to the total VDOS we also calculate the partial vibrational density of states for O, Si, and Na (including subdivisions based on $Q^n$ distribution and oxygen coordination, see Figure~S15 in the Supplementary Material as well as an example of the decomposed VDOS in Figure~S16). From this analysis we find that the shift of the low-frequency band (0-10~THz) in Figure~\ref{fig2_vdos} to higher frequencies and intensity is caused by sodium contributions while the shifting in the 10-25~THz range are found to be caused by oxygen contributions upon increasing Na$_2$O content.

We further show the eigenvectors (i.e., the solutions to Eq.~\ref{eq.eigenvectors}) of three modes in Figure~\ref{fig2_vdos} to graphically illustrate the vast changes of modal characteristics in different frequency ranges. For example, in Figure~\ref{fig2_vdos}a a low frequency mode at $\sim$7~THz is illustrated with a modal phase quotient (PQ) value (PQ=0.77), i.e. a mode where eigenvector displacements are mostly in-phase (vibrating in the same direction). Interestingly, this agrees with previous findings that modes in phase are common in the low-frequency region in both glasses and crystalline solids~\cite{Seyf2018}, which could suggest that modes with high a PQ resemble that of a classical phonon. Observing the mid-frequency range ($\sim$10-30~THz) where modes tend to shift towards diffusive heat transfer processes (often denoted as diffusons), we illustrate a mode at 20~THz in Figure~\ref{fig2_vdos}b. Compared to Figure~\ref{fig2_vdos}a we find that displacements seem directionally random yet that vibrations are still extended throughout the solid in the majority of atoms. Finally, we inspect a high-frequency mode (32 THz) in Figure~\ref{fig2_vdos}c which shows highly localised eigendisplacements with vibrations centered around only few atoms. These modes are often referred to as locons which are commonly found at higher frequencies above the so-called 'mobility edge' (generally, modes of participation ratios lower than 0.15)~\cite{Allen1989}. To probe the effect of the depolymerization of the network with increasing Na$_2$O content we provide a more elaborate illustration of eigenvectors in multiple compositions at a series of frequencies in Figure~S17 in the Supplementary Material. Notably, it is found that eigenvectors in modes with a frequency of $\sim$1~THz become less periodic and more diffuse when increasing the sodium content - a feature often related to decreasing heat transfer~\cite{Zhou2019,Agne2018}. Furthermore, eigenvectors in modes with a frequency of 15~THz in the mid-frequency range seems to become more localized in certain spatial volumes when increasing the sodium concentration, which is an indication of lowered participation ratio of eigenvectors when disrupting the silica network. Finally, in the high-frequency range at 35~THz, modes are generally very localized in all simulated glasses, yet upon increasing Na$_2$O content, eigenvectors seem to gradually shift towards higher localization.

\begin{figure}[!htb]
    \centering
    \includegraphics[width = 8.6cm]{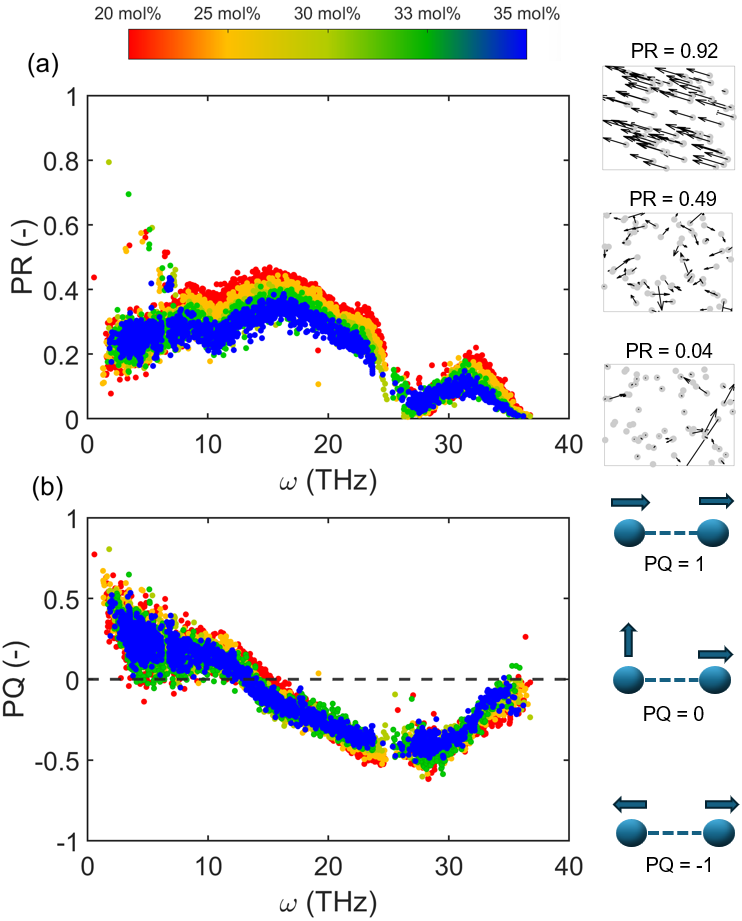}
    \caption{(a) Participation ratio (PR) of all studied sodium silicate glasses (right side of a) along with three modes illustrating the connection between modal character and PR; and (b) phase quotient (PQ) as well as an illustration of the relation between nearest neighbor eigenvectors at PQ~=\{-1, 0, 1\} for all studied $x\text{Na}_2\text{O}\text{-}(100\text{-}x)\text{SiO}_2$ glasses with increasing sodium contents.}
    \label{fig4_PR_PQ}
\end{figure}

While the visual inspection of eigenvectors is useful to provide an overview, to quantitatively assess the qualitative observations of mode localization provided above, we compute the participation ratio as presented in Figure~\ref{fig4_PR_PQ}a. Here, we observe that mid-frequency (10-20~THz) modes possess the highest PR ($\sim$0.4), suggesting that the atomic eigenvectors in modes similar to the one illustrated in Fig~\ref{fig2_vdos}b are somewhat the same magnitude and therefore contribute evenly to the vibrational motion in diffusive processes. Upon increasing the sodium concentration we observe how the PR decreases in the mid-frequency range and shifted towards an increase in the low-frequency range. This observation coincides with the shift of the partial VDOS of sodium and bridging oxygen species (See Figure~S15 in the Supplementary Material), where the contribution from sodium related vibrations increase in the low frequency range while the contribution from bridging oxygen species decrease in the 10-23~THz range. The partial VDOS of silicon show that the atomic contribution from $Q^4$ and $Q^3$ silicon species also decreases at $\sim$20~THz, likely due to the induced network depolymerization, suggesting that sodium atoms in the interstitial sites are responsible for localization changes of atomic vibrations in the silica backbone.

Additionally, from Figure~\ref{fig2_vdos} we find that high-frequency modes (>27~THz) are slightly shifted towards lower frequencies upon increasing Na$_2$O content. From Figure~\ref{fig4_PR_PQ}a and Figure~S18 in the Supplementary Material we find that this shift is accompanied by a slight decrease in $PR$. Generally, modes <27~THz have a $PR>0.15$ and are thus non-localized while while modes >27~THz are localized ($PR<0.15$), albeit we note that localization is rather a gradual than a discrete change. 

To obtain further insight, we compute the atomic participation ratio ($APR_{\alpha}$ where $\alpha$ denotes an atomic type) in Figure~S19 in the Supplementary Material, which provides a neat way of describing contributions of different contributions to $PR$, in a way, such that $PR(n)$= $\sum_{i\in\alpha}$ ($APR_{\alpha}(n))^2$. 

Here we find that the increased participation ratio in the low-frequency range ($\sim$0-8~THz) are in fact a consequence of decreased localization of sodium vibrations. These localization changes quite impactfully increase the localization of bridging-oxygen vibrations in the range $\sim$0-20~THz, suggesting that sodium has much higher impact on oxygen vibrations than silicon-related vibrations, which seems to be reasonable observation since sodium is coordinating only directly to oxygens in the present systems and therefore has only weak interactions with silicon.

Due to their amorphous nature, dispersion relations are not well-described in glasses~\cite{greaves1985exafs}. Instead, to access the acoustical/optical character of modes we compute the so-called phase quotient ($PQ$) per mode as introduced by Bell and Hibbins-Butler~\cite{Bell1975}. We present the $PQ$ in Fig~\ref{fig4_PR_PQ}b, illustrating three distinct regions; i.e. modes vibrating in-phase~(i.e., eigenvectors of neighboring atoms have angles of <90$^\circ$, and thus $PQ>0$, see Methods for definition); modes where nearby atoms vibrate perpendicular to each nearest neighbor eigenvector~($PQ=0$); and finally modes out of phase~(angle between eigenvectors >90$^\circ$ and $PQ<0$). We find that the majority of low-frequency  modes are in-phase vibrations, which is a common feature in both crystalline and amorphous solids~\cite{Seyf2018,Bell1975, Voyles2001}. 

Here, the PQ of the simulated glasses transitions from in-phase to out of phase at $\sim$13~THz and then decreases until about $\sim$25~THz. Modes above this range is found to increase in $PQ$, but appears to level off at $PQ\sim0$. A plot of the $PQ$ of a 3000 atom system is shown in Figure~S20 in the Supplemental Material, showing that high-frequency out-of-phase modes vary considerably in $PQ$ 

We already established that depolymerization of the silica network with Na$_2$O addition promotes considerable changes in modal characteristics of low and mid-frequency modes ($\sim$0-25~THz), e.g. disruption of somewhat periodic eigenvectors at 1~THz (based on qualitative observations, see Figure~S17 in the Supplementary Material) and decreased localization of modes in the mid-frequency range ($\sim$8-25~THz) as a consequence of increased modal density of sodium related vibrations and extended localization of modes in the low-frequency range ($\sim$0-8~THz). When observing the $PQ$ in the mentioned frequency ranges, we find that the $PQ$ of modes converge towards more perpendicular (or random) vibrations, i.e. $PQ~\rightarrow$~0, besides a small range of modes with frequencies of $\sim$14-18~THz. The most notable changes are found in the low-frequency range, where the $PQ$ of modes on average decrease, which correlates with the observations found in the modal illustrations in Figure~S17 where eigenvectors seem to become more diffusive. In addition to $PQ$, we compute the atomic phase quotient ($APQ$, see Methods) of Si, O, and Na as a function of modal frequency (see Figure~S21). We note that the total $PQ$ is not the sum of the $APQs$. Rather, the $APQ$ provides a description of the phase-characteristics on a atom-type basis which is an interesting way of distinguishing the local vibrational environments.

Generally, we observe that the majority of changes are related to silicon and oxygen vibrations, where silicon seems to become more in-phase with adjacent atoms while oxygen vibrations become less in phase with adjacent atoms when increasing the sodium content. This might be due to the increased order in the tetrahedral environments of silicon as observed in the O-Si-O angles provided in Figure~S6 in the Supplementary Material. Another interesting observation is how the Na $APQ$ is near 0, indicating a significant degree of randomicity in the vibrational directions compared to the nearest neighbors, likely due to the weak bonding environment around Na atoms - a feature which is becoming more pronounced with a higher amount of Na$_2$O in the glass.

\subsection{Heat transfer}

\begin{figure}[!t]
    \centering
    \includegraphics[width = 0.9\columnwidth]{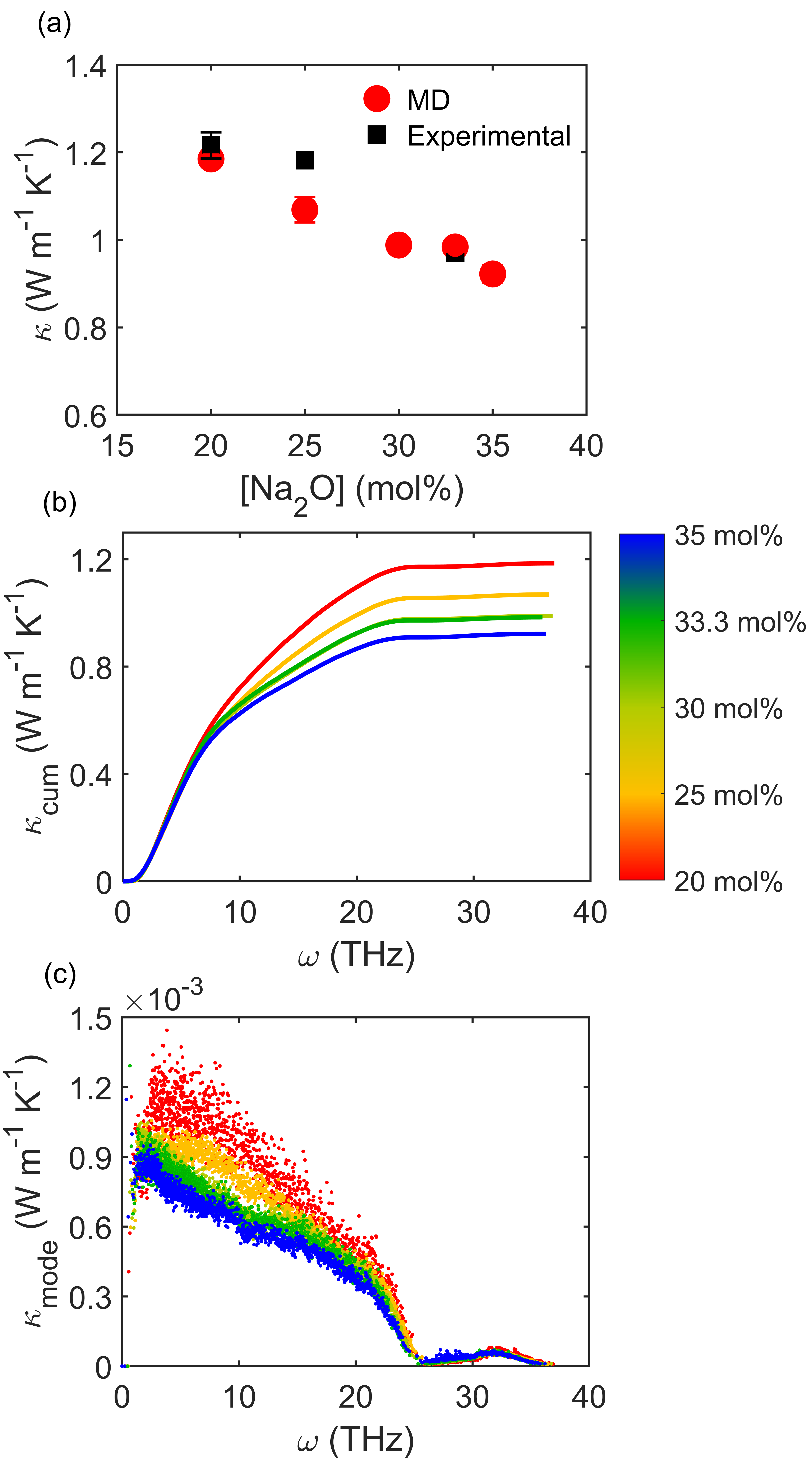}
    \caption{Effect of sodium concentration on (a) total thermal conductivity (b), cumulative thermal conductivity and (c) modal thermal conductivity of simulated $x\text{Na}_2\text{O}\text{-}(100\text{-}x)\text{SiO}_2$ glasses. Experimental values are from Ref.~\cite{Hiroshima2008,Sukenaga2021}. Error bars for MD glasses are smaller than the symbols.}
    \label{fig3_cond_cumcond}
\end{figure}

Despite some interest in the fundamental vibrational characteristics of a few of the sodium silicate glasses\cite{Kilymis2019,bauchy}, derived properties, e.g. thermal conduction, are practically unexplored in simulations. This lack of study is notable given the well-understood structure and mechanical properties of these glasses as well as their usefulness as an archetypical model system for modified alkali silicate glasses, as well as its similarity to many glasses used industrially (e.g., windows and screens)~\cite{Varshneya2013}. In this work we employ the recently developed method of estimating the thermal conductivity ($\kappa$) in a unified model~\cite{Isaeva2019,Simoncelli2019} coined as the Quasi Harmonic Green Kubo (QHGK)~\cite{Isaeva2019}. This model allows the estimation of $\kappa$ beyond the pure phonon, diffuson (i.e. non-localized modes which see short relaxation times) and locon (i.e. localized modes which see only vibration of few atoms and hence negligible contribution to $\kappa$) regimes. Practically, the model requires the input of force constants to the third order. We computed these through a finite displacement method implemented in LAMMPS and performed the following QHGK calculations as implemented in kALDo~\cite{Barbalinardo2020}. We present the total thermal conductivity estimations in Figure~\ref{fig3_cond_cumcond}a where we also compare with experimental values of thermal conductivity~\cite{Hiroshima2008,Sukenaga2021}. 

Generally, we find very good agreement between simulated and experimental data in the observed range, which seems to be consistent with the well reproduced VDOS of simulated glasses (Figure~\ref{fig2_vdos}). Notably, the glasses with molar compositions $x$~=~20 and $x$~=~33.3 deviate from experimental data with 2.6 and 1.3\%, respectively, showcasing how the Teter potential is able to predict vibrational features and thermal conductivity nicely. This agreement is a significant finding given how MD simulations have been notoriously known for overestimating $\kappa$ of many glass systems~\cite{chojin2020cooling,Soerensen2019}. Practically, the estimations show a linear decrease in $\kappa$ from $\sim$1.2 to $\sim$0.9~W~m$^{-1}$~K$^{-1}$ when increasing the sodium concentration from 20 to 35 mol\%, evidently showing that formation of non-bridging oxygen from network depolymerization results in linearly decreasing $\kappa$ in the simulated glasses.

A clear strength of QHGK calculations is that it does not only provide the total $\kappa$ but also an individual contribution of each mode to the total $\kappa$, a so-called 'modal' thermal conductivity. From this, we present the cumulative thermal conductivity in Figure~\ref{fig3_cond_cumcond}b to investigate the contribution to the thermal conductivity ($\kappa_{\text{cum}}$) in different frequency ranges upon increasing the sodium concentration. Upon increasing Na$_2$O contents we find that the contribution to $\kappa$ from low-frequency modes (0-8~THz) remain somewhat unchanged, while the slope of the curves decrease in the mid-frequency range (8-24~THz), i.e. the contribution to $\kappa$ of modes in the mid-frequency range decreases. 

We furthermore present the modal contribution to the thermal conductivity~($\kappa_{\text{mode}}$) in Figure~\ref{fig3_cond_cumcond}c to investigate the contribution to $\kappa$ from each of the 3$N$ modes individually. From this plot, it is evident that lower frequency modes contribute the most to $\kappa$, and the modal contribution seems to decrease with increasing frequency. Notably, above 25~THz, modes exhibit very low contribution to $\kappa$ in line with their very low PR (Figure~\ref{fig4_PR_PQ}a)~\cite{Allen1989}. For more, we find a significant decrease in $\kappa_{\text{mode}}$ values in the frequency range 0-8~THz when increasing the sodium concentration, while the number of modes increases as seen from the increased modal density in the VDOS (Figure~\ref{fig2_vdos}), which explains the somewhat static $\kappa_{\text{cum}}$ in the low-frequency range (Figure~\ref{fig3_cond_cumcond}b). Notably, the increased modal density of the VDOS in the range 0-8~THz originates from increased vibrational contributions from sodium related vibrations as seen in the partial VDOS of sodium (See Figure~S15 in the Supplementary Material), suggesting that sodium related vibrations are responsible for a decreased $\kappa$ contribution in the lower frequency modes. In addition to values of $\kappa$, the modal diffusivities $D_i$ from the QHGK calculations are shown in Figure~S22 in the Supplementary Material illustrating that the diffusivity of modes significantly decrease in the 0-25~THz frequency range upon increasing Na$_2$O. However, we find that the diffusivity of modes feature a rather constant contribution across the frequency range. This difference in comparison to thermal conductivity (which see a decreasing contribution, see Figure~\ref{fig3_cond_cumcond}c) is caused by the quantum effect of heat capacity with decreasing mode occupancy levels at higher frequencies.

\subsection{Correlating modal characteristics and heat transfer}

\begin{figure}[!b]
    \centering
    \includegraphics[width = 9cm]{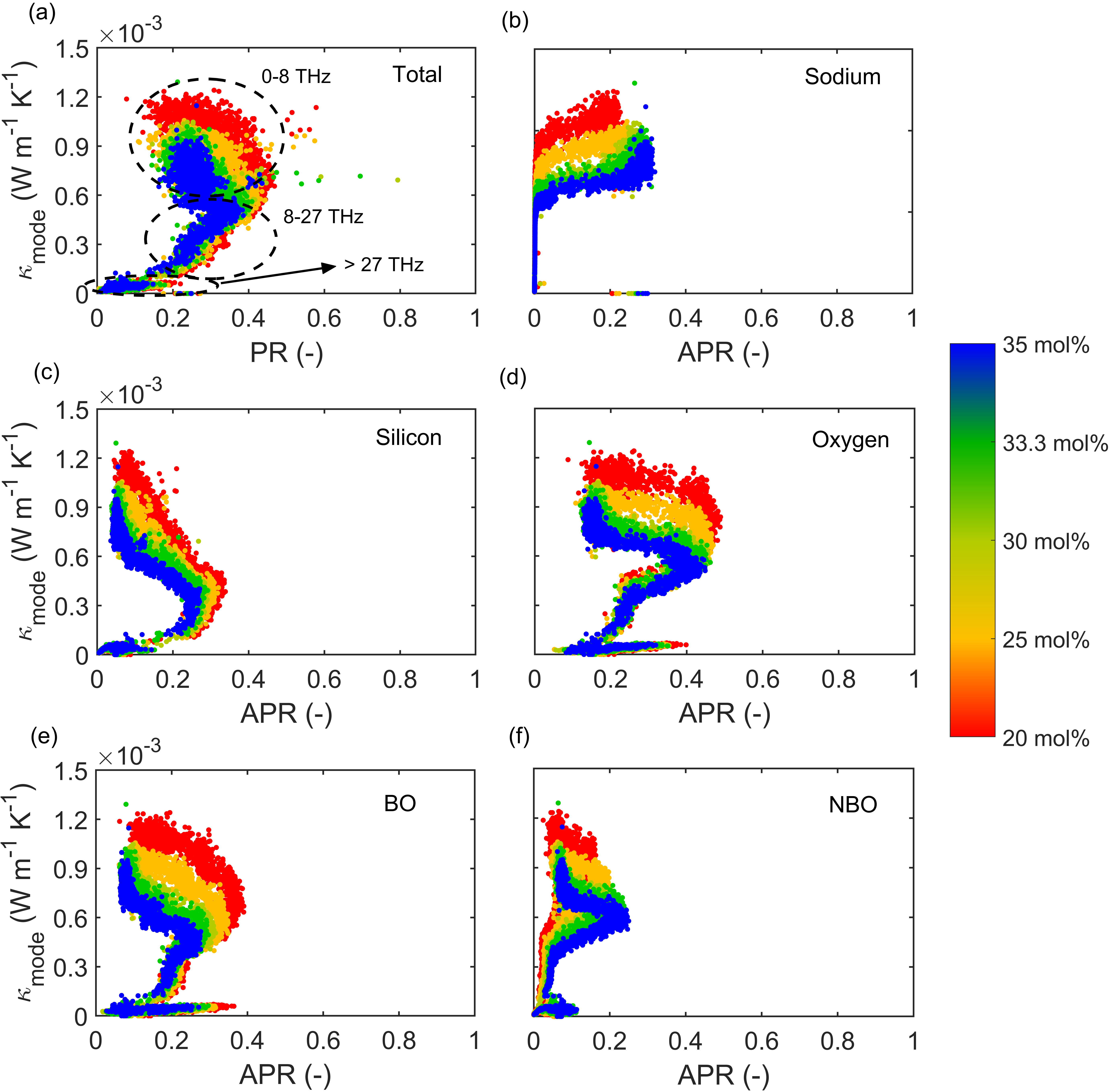}
    \caption{Effect of sodium concentration on (a) total participation ratio and atomic participation ratio of (b) sodium, (c) silicon, (d) oxygen, (e) bridging-oxygens and (f) non-bridging oxygens coupled with modal thermal conductivities estimated with the QHGK method.}
    \label{fig5_cond_PR_APR}
\end{figure}

Next, to provide additional insight to the heat transfer mechanisms of the studied glasses, the modal thermal conductivities provided in Figure~\ref{fig3_cond_cumcond}c are coupled with the computed vibrational properties of $PR$ and $APQ$ (Figure~\ref{fig4_PR_PQ} as well as Figures~S19-21). In Figure~\ref{fig5_cond_PR_APR}a we compare the modal participation ratio of glasses with the modal thermal conductivity and study the effect of localization changes upon increasing the Na$_2$O concentration. 

We first notice the small band in the PR range $\sim$0-0.2 where modes are contributing insignificantly to $\kappa$ (i.e., $\kappa_{\text{mode}}<0.1\cdot 10^{-3}$~W~m$^{-1}$~K$^{-1}$). These are high frequency modes (>27~THz) which see a high spatial localization, which is consistent with previous findings of other amorphous systems in the literature~\cite{Allen1989,Lv2016}. For modes found in the frequency range 8-27~THz, we find that $\kappa_{\text{mode}}$ increases in a linear fashion with increasing $PR$. Increasing the $PR$ of modes could rationally lead to the assumption that the modal contribution to $\kappa$ increases as a consequence of more extended energy propagation. However, for modes found in the frequency range 0-8~THz, the modal contribution to $\kappa$ seems to be inversely proportional to $PR$, which is in disagreement with the trend found for modes in the range of 8-27~THz. These observations show that the correlation between the modal participation ratio and the per-mode contribution to $\kappa$ of extended modes is not straightforward in the studied sodium silicate glasses, thus suggesting that $PR$ may not be the ideal descriptor of the underlying heat transfer.

In Figures~\ref{fig5_cond_PR_APR}b-f, we coupled $\kappa_{\text{mode}}$ with the APR of different atomic species (Si, Na, O, as well as the decompositions into bridging and non-bridging oxygens). We first turn our attention to the APR of sodium atoms coupled with $\kappa_{\text{mode}}$ in Figure~\ref{fig5_cond_PR_APR}b. As previously assessed, the modal participation ratio increases in the low-frequency range when increasing the sodium concentration (Figure~\ref{fig4_PR_PQ}), which results from a decreased vibrational localization of sodium atoms. Furthermore,  $\kappa_{\text{mode}}$ decreases as the localization of sodium related vibrations decreases, yet, above APR$\sim$0.22 the $\kappa_{\text{mode}}$ gradually increases in each glass with more than 25~mol\% of $\text{Na}_2\text{O}$, which could be correlated to a greater heat transfer contributed from more evenly distributed sodium vibrations. Since higher degree of polymerization in the silica backbone (and thus less Na$_2$O) has shown to be associated with greater heat transfer, we now study the localization of silion and oxygen to investigate how disruption of the silica network impacts localization of the network forming atoms and ultimately $\kappa$. 

In Figure~\ref{fig5_cond_PR_APR}c the $\kappa_{\text{mode}}$ coupled with $APR$ of silicon is presented. Similar to the total $PR$ (Figure~\ref{fig5_cond_PR_APR}a), a clear correlation between localization of silicon and $\kappa_{\text{mode}}$ is difficult to establish. Although, we observe that an increasing Na$_2$O content increases the localization of silicon in almost the entire spectral range, which seems to correlate with a decrease in $\kappa_{\text{mode}}$. In general, the silicon vibrations contribute very little to the modes of high $\kappa_{\text{mode}}$ as opposed to sodium, suggesting that the localization changes of silicon may have less impact on overall thermal conduction. Lastly, we showcase the correlation between oxygen $APR$ and $\kappa_{\text{mode}}$ in Figure~\ref{fig5_cond_PR_APR}c-f. We find that oxygen in general show similar trends to that of the total $PR$, especially in glasses with lower sodium contents. In fact, when observing the $APR$ of bridging oxygens presented in Figure~\ref{fig5_cond_PR_APR}d-f, we find that oxygen in the silica backbone possess higher atomic participation ratio than NBOs and silicon atoms as seen in Figure~\ref{fig5_cond_PR_APR}c,e,f
, suggesting BOs to be the main contributor to the heat transfer in the mid-low frequency range ($\sim$8-20~THz) in the studied glasses. As might be noted, the $APR$ is dependent on the number of atoms in each sample. As such, the changing composition also inherently changes the $APR$. To make a more unbiased comparison we also computed an $APR$ normalized by the number of the atom type in each composition and correlated this to $\kappa_{\text{mode}}$. We present this plot in Figure~S23 in the Supplementary Material showcasing how the range of the normalized $APR$ is largely similar for the different atomic types across the varying compositions. This indicates a simple scaling of the $APR$ with the composition change. In summary, while there seems to be some correlations between thermal conductivity and $PR$, the correlation is nontrivial.

\begin{figure}[!b]
    \centering
    \includegraphics[width = 8.6cm]{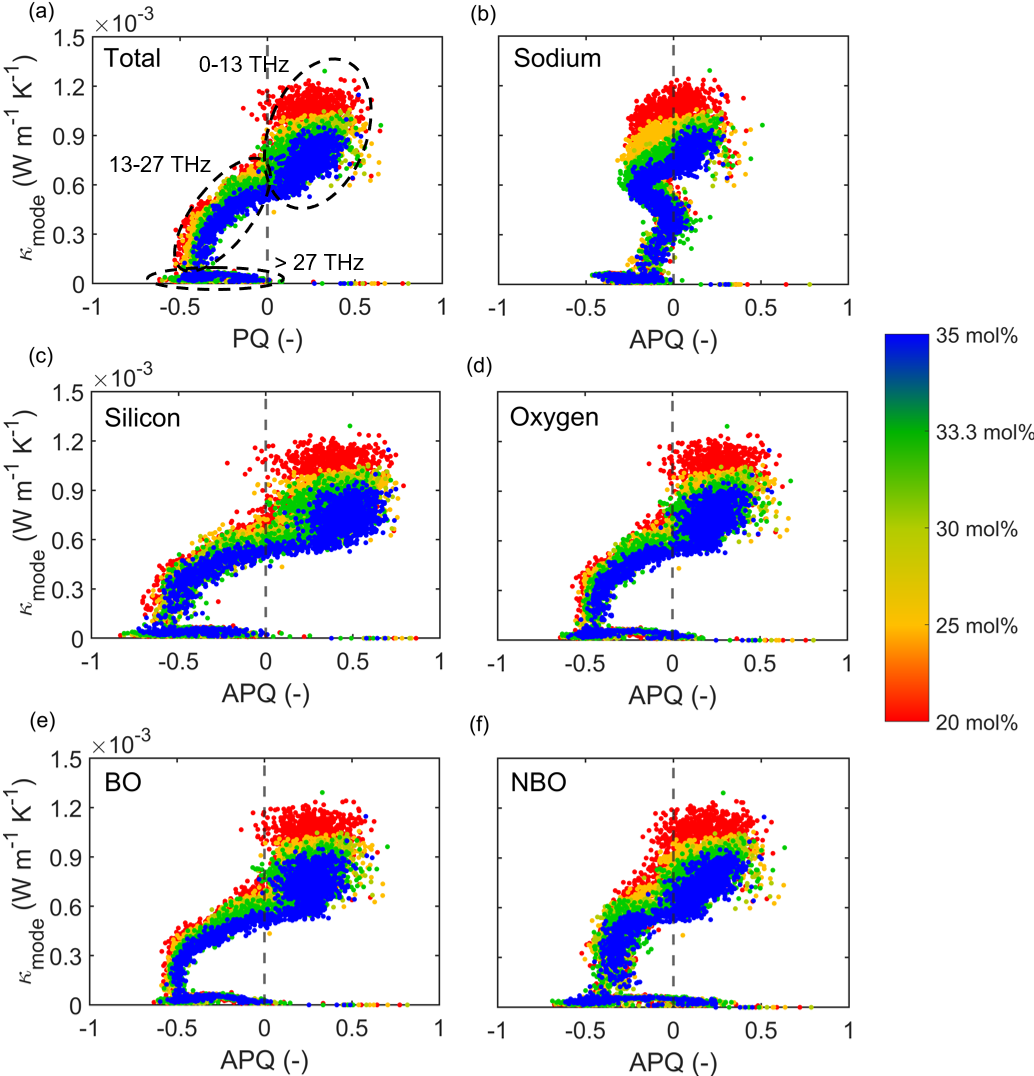}
    \caption{ $\kappa_{\text{mode}}$ plotted as a function of (a) the total phase quotient (PQ) of all nearest neighbor atomic interactions in each mode, as well as the atomic phase quotient (APQ) providing the angle between eigenvectors of nearest neighbor atoms of (b) sodium, (c) silicon, (d) oxygen (e) bridging-oxygens, and (e) non-bridging oxygens, respectively, for each of the studied sodium silicate glasses. Dashed lines provide the transition point (PQ=0) from in-phase to out of phase modes.}    
    \label{fig6_cond_PQ}
\end{figure}

Instead, we investigate the impact of in-phase and out-of-phase modes, through the use of the phase-quotient ($PQ$), and its relation to the thermal conductivity. We provide these correlations for the total system as well as for all atom types in Figure~\ref{fig6_cond_PQ}. We generally observe higher $PQ$ (i.e., more in-phase vibrations) at lower frequencies (Figure~\ref{fig4_PR_PQ}) which is somewhat similar to many crystalline solids~\cite{Seyf2018}. For more, we generally find that higher $PQ$ is associated with higher modal $\kappa$ (Figure~\ref{fig6_cond_PQ}). This may be correlated to the inherent periodicity of strongly acoustic phonons found in crystalline solids which are also known as strong contributors to thermal conductivity. This suggests that the in-phase modes observed in the sodium silicate glasses possess some degree of periodicity, and hence explain the relatively high contribution to $\kappa$. With increasing Na$_2$O content we generally observe small changes to total $PQ$ (Figures~\ref{fig4_PR_PQ} and \ref{fig6_cond_PQ}), however a slight decrease in the associated $\kappa_{\text{mode}}$ is found. For more, a broad band can be found in the range of $PQ=-0.5$ to $PQ=0$ with very low contribution to $\kappa$ corresponding to the localized modes observed in Figure~\ref{fig4_PR_PQ}a. This indicates that localized modes with low contribution to the thermal conductivity inherently consist of out-of-phase vibrations of nearest neighbour atoms.

For more, from Figure~\ref{fig4_PR_PQ} we find that the $PQ$ becomes less spread in the low-frequency range, while also shifting slightly towards $PQ~=~0$ (i.e., they loose their character as either acoustic or optical) in the whole spectral range when the sodium concentration is increased, indicating a sort of disordering of the vibrations. To study these features and their impact on the modal contribution to $\kappa$, we investigate the atomic phase quotient (APQ), which describes the angle  between specific atomic types (O, Si, Na, bridging-, and non-bridging O) and the eigenvector of their nearest neighbour atoms (see illustration in Figure~\ref{fig4_PR_PQ}b). The APQs are plotted in Figure S21 in the Supplementary Material as a function of modal frequencies, showing some of the same trends as the total $PQ$ shown in Figure~\ref{fig4_PR_PQ}. We now couple the $PQ$ and APQs to $\kappa_{\text{mode}}$ (Figure~\ref{fig6_cond_PQ}) to investigate the impact of increasing the Na$_2$O on nearest neighbor atomic vibrations and the thermal conductivity. 

The silicon and oxygen APQs in Figure~\ref{fig6_cond_PQ}(c,d) show many similarities to the trends observed in the total PQ. For example, there seems to be a very large degree of similarity in plots in Figure~\ref{fig6_cond_PQ} when changing the molar concentration of Na$_2$O. Effectively, this means that modes containing similar oxygen and silicon vibrational characteristics contribute largely similarly to heat conduction, independent of the glass composition. The small decrease in $\kappa_{\text{mode}}$ for vibrations of otherwise similar PQ/APQ is likely caused by the disruptive nature of the Na$_2$O addition. Interestingly, for oxygen these points seem to be gradually different for BOs and NBOs (Figure~\ref{fig6_cond_PQ}(e,f)), namely that the BOs seem to provide modes of stronger acoustic and optical character than NBOs. This points towards a slightly higher "disorder" in the vibrations of the NBOs, a less distinct correlation with $\kappa_{\text{mode}}$ and hence likely a smaller direct correlation with heat transfer.

 Similarly , when observing the sodium APQ provided in Figure~\ref{fig6_cond_PQ}b, we observe a rather distinct APQ trend. Although the correlation between APQ and $\kappa_{\text{mode}}$ is not easily defined, we observe that modes generally lie closer to PQ=0 than the silicon and oxygen APQs. This may be linked to how sodium atoms are more loosely bound in the glass structure, effectively causing the vibrational direction in each mode to be less coupled to the direction of vibration of its nearest neighbor (which is always an oxygen, most commonly in an octahedral state, see Table~S1 in the Supplementary Material). Effectively, this is believed to disrupt effective heat transfer throughout the network despite its large contribution to lower frequency vibrations, effectively reducing the per-mode thermal conductivity contribution (Figures~\ref{fig4_PR_PQ}-\ref{fig6_cond_PQ}). Naturally, upon increasing contents of Na$_2$O, this effect is enhanced, explaining the overall reduction in $\kappa$ (Figure~\ref{fig3_cond_cumcond}) in modified oxide glass systems.

\section{Conclusions}
In summary, using classical molecular dynamics simulations we studied a series of sodium silicate ($x\text{Na}_2\text{O}\text{-}(100\text{-}x)\text{SiO}_2$) glasses with varying Na$_2$O content ($x=20-35$~mol\%) using the recently developed quasi-harmonic Green Kubo method. We find the used classical potential to be capable of predicting the observed experimental thermal conductivity of the systems and the decreasing thermal conductivity with increasing Na$_2$O content. By modal deconvolution of the contributions to the total thermal conductivity, we find the per-mode contribution to conductivity to generally decrease in a wide frequency span (0-25~THz) upon increasing Na$_2$O content, however due to a shifting of modes from 8-25~THz to 0-8~THz, the overall contribution to thermal conductivity is solely decreasing in the 8-22~THz frequency range. 

From a detailed vibrational analysis we identify a clear trend of increasing thermal conductivity with increasing in-phase motion of the underlying vibrations, especially for the network forming atoms (Si and O). However, upon increasing Na$_2$O concentrations we find atomic vibrations to become less 'ordered', effectively causing a disruption of heat transfer throughout the glass network. Ultimately, our results provide new insight to the connection between vibrational character and mode-resolved thermal conductivity in the family of modified oxide glasses and may aid in the future design of oxide glasses of tailored thermal properties. 

\section*{SUPPLEMENTARY MATERIAL}
See Supplemental Material for tabulated values of structural, mechanical, and thermal properties as well as detailed plots of pair-potential, density, pair-distribution functions, bond-angle distributions, structure factors, vibrational density of states, eigenmodes, and various vibrational characteristics of the studied glass systems.

\section*{Author Declarations}
\subsection*{Conflicts of interest}
The authors declare that they have no competing interests.

\section*{Acknowledgments}
The authors acknowledge computational resources provided by CLAAUDIA at Aalborg University. The present work was initiated as a student thesis project and we acknowledge the thesis opponent Nicholas Bailey (Roskilde University) for fruitful discussions.

\section*{Author contributions}
\textbf{Philip Rasmussen}: Conceptualization (supporting); Methodology (equal); Investigation (lead); Formal analysis (lead); Visualization (lead); Writing – original draft (lead); Writing - review and editing (supporting).  \textbf{Søren S. Sørensen}: Conceptualization (lead); Methodology (equal); Formal analysis (supporting); Supervision (lead); Writing – original draft (supporting); Writing - review and editing (lead).

\section*{Data Availability Statement}

The data that support the findings of this study are available from
the corresponding author upon reasonable request.

\bibliography{aipsamp}

\newpage

\onecolumngrid



\section*{Supplementary material}

# SUPPLEMENTARY MATERIAL
# Thermal conductivity in modified oxide glasses is governed by modal phase changes

P. Rasmussen[1] and S. S. Sørensen[1]

1    Department of Chemistry and Bioscience, Aalborg University, Aalborg, Denmark, |
soe@bio.aau.dk 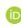 https://orcid.org/0000-0003-2230-7823

Table S 1: Molar concentration of $Na_2O$ ($x$) in the $x$$Na_2O$-(100-$x$)$SiO_2$ glasses, mass density ($\rho_m$), atomic density ($\rho_A$), average molar mass ($M_{avg}$), simulation cell size, $l_{cell}$ and number of atoms in each simulation cell used in simulations containing 3000 atom box sizes. Mass densities are adapted from Refs. [1], [2].

| $x$ (mol%) | $\rho_m$ (g cm$^{-3}$) | $\rho_A$ (Å$^{-3} \cdot 10^{-2}$) | $M_{avg}$ (g mol$^{-1}$) | $l_{cell}$ (Å) | # of atoms in simulation cell | | |
|---|---|---|---|---|---|---|---|
| | | | | | Na | O | Si |
| 0 | 2.2 | 6.62 | 20.028 | 35.66 | 0 | 2000 | 1000 |
| 10 | 2.289 | 6.86 | 22.091 | 35.23 | 200 | 1900 | 900 |
| 15 | 2.336 | 6.99 | 20.123 | 35.01 | 300 | 1850 | 850 |
| 20 | 2.383 | 7.17 | 20.154 | 34.80 | 400 | 1800 | 800 |
| 25 | 2.431 | 7.25 | 20.186 | 34.58 | 500 | 1750 | 750 |
| 30 | 2.466 | 7.35 | 20.217 | 34.44 | 600 | 1700 | 700 |
| 33.3 | 2.491 | 7.41 | 20.238 | 34.33 | 666 | 1667 | 667 |
| 35 | 2.497 | 7.43 | 20.249 | 34.31 | 700 | 1650 | 650 |
| 40 | 2.532 | 7.52 | 20.28 | 34.17 | 800 | 1600 | 600 |
| 45 | 2.541 | 7.53 | 20.312 | 34.15 | 900 | 1550 | 550 |
| 50 | 2.560 | 7.58 | 20.344 | 34.01 | 1000 | 1500 | 500 |

Table S 2: Molar concentration of $Na_2O$ ($x$) in the $x$$Na_2O$-(100-$x$)$SiO_2$ glasses, mass density ($\rho_m$), atomic density ($\rho_A$), average molar mass ($M_{avg}$), simulation cell size, $l_{cell}$ and number of atoms in simulations used for estimating thermal properties with the Quasi-harmonic Green-Kubo approach (600 atoms). Mass densities are adapted from Refs. [1], [2].

| $x$ (mol%) | $\rho_m$ (g cm$^{-3}$) | $\rho_A$ (Å$^{-3}$·10$^{-2}$) | $M_{avg}$ (g mol$^{-1}$) | $l_{cell}$ (Å) | # of atoms in simulation cell | | |
|---|---|---|---|---|---|---|---|
| | | | | | Na | O | Si |
| 20 | 2.383 | 7.17 | 20.154 | 20.35 | 80 | 360 | 160 |
| 25 | 2.431 | 7.25 | 20.186 | 20.22 | 100 | 350 | 150 |
| 30 | 2.466 | 7.35 | 20.217 | 20.14 | 120 | 340 | 140 |
| 33.3 | 2.491 | 7.41 | 20.238 | 20.08 | 132 | 335 | 133 |
| 35 | 2.497 | 7.43 | 20.249 | 20.07 | 140 | 330 | 130 |

Table S 3: Average bond lengths ($l$) of Si-O and Na-Ointeractions , and coordination numbers (CN) of Na and Si of simulated sodium silicate glasses at various sodium concentrations ($x$). Experimental data is provided as extracted from neutron diffraction studies. References are provided with the experimental values.

| $x$ (mol%) | $l_{Si-O}$ (Å) | $l_{Na-O}$ (Å) | $l_{Si-O,exp}$ (Å) | $l_{Na-O,exp}$ (Å) | $CN_{Na}$ (−) | $CN_{Si}$ (−) | $CN_{Si-O,exp}$ (−) | $CN_{Na-O,exp}$ (−) |
|---|---|---|---|---|---|---|---|---|
| 0 | 1.6125 | | 1.610 | | | 4.0 | 4.0 [3] | |
| 10 | 1.6125 | 2.3625 | | | 5.27 | 4.0 | | |
| 15 | 1.6125 | 2.3775 | | | 5.86 | 4.0 | | |
| 20 | 1.6125 | 2.3925 | 1.617 | 2.36 | 6.12 | 4.0 | | |
| 25 | 1.6125 | 2.3775 | | | 6.09 | 4.0 | | 5 [4] |
| 30 | 1.6125 | 2.3775 | 1.622 | 2.36 | 5.89 | 4.0 | | |
| 33 | 1.6125 | 2.3775 | | | 5.79 | 4.0 | 4.0 [3] | 5.5 [3] |
| 35 | 1.6125 | 2.3775 | | | 6.08 | 4.0 | | |
| 40 | 1.6125 | 2.3775 | | | 6.14 | 4.0 | | |
| 45 | 1.5975 | 2.3775 | | | 6.06 | 4.0 | | |
| 50 | 1.5975 | 2.3775 | | | 5.98 | 4.0 | | 5 [4] |

Table S 4: Elastic properties of glasses investigated, where $x$ is the molar concentration of $Na_2O$ ($x$) in the $x Na_2O$-(100-$x$)$SiO_2$ glasses, $E$ is the Young's modulus, $B$ is the bulk modulus, $G$ is the shear modulus and $v$ is the Poisson's ratio. Data provided are estimated from 3000 atom system sizes.

| $x$ (mol%) | $E$ (GPa) | $B$ (GPa) | $G$ (GPa) | $v$ (-) |
|---|---|---|---|---|
| 0 | 101.3 | 65.5 | 40.8 | 0.2415 |
| 10 | 79.2 | 51.2 | 31.9 | 0.2422 |
| 15 | 72.2 | 47.1 | 29.0 | 0.2441 |
| 20 | 66.8 | 43.6 | 26.8 | 0.2441 |
| 25 | 63.0 | 40.6 | 25.4 | 0.2402 |
| 30 | 56.9 | 38.8 | 22.7 | 0.2553 |
| 33 | 56.2 | 39.9 | 22.3 | 0.2637 |
| 35 | 53.9 | 40.2 | 21.1 | 0.2756 |
| 40 | 52.3 | 39.7 | 20.4 | 0.2794 |
| 45 | 49.1 | 37.4 | 19.2 | 0.2811 |
| 50 | 45.7 | 39.4 | 17.5 | 0.3053 |

Table S 5: Thermal conductivities data computed with the Quasi-Harmonic Green-Kubo method of $x$Na$_2$O-(100-$x$)SiO$_2$ glasses where ($x$) is the molar concentraion of Na$_2$O, ($\kappa$) is the average thermal conductivity and (SD) is the standard deviations of estimated $\kappa$ values.

| $x$ (mol%) | $\kappa$ (W m$^{-1}$ K$^{-1}$) | SD (W m$^{-1}$ K$^{-1}$) |
|---|---|---|
| 20 | 1.185 | 0.006 |
| 25 | 1.069 | 0.029 |
| 30 | 0.986 | 0.015 |
| 33 | 0.986 | 0.012 |
| 35 | 0.915 | 0.007 |

**Figure S1**

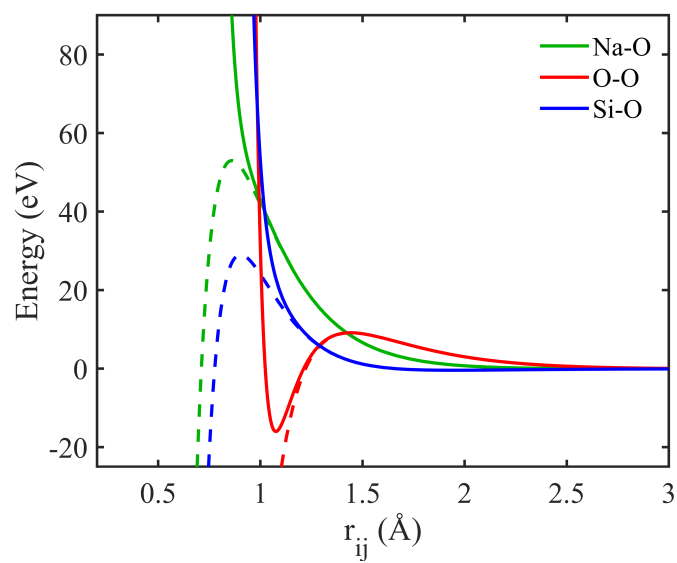

Figure S 1: Potential energy curve for the Na-O, O-O and Si-O interactions of the Teter potential plotted without the $r^{24}$ short term correction (dashed lines) and with the short term correction (solid lines), illustrating the modification added to avoid the "Buckingham catastrophe".

**Figure S2**

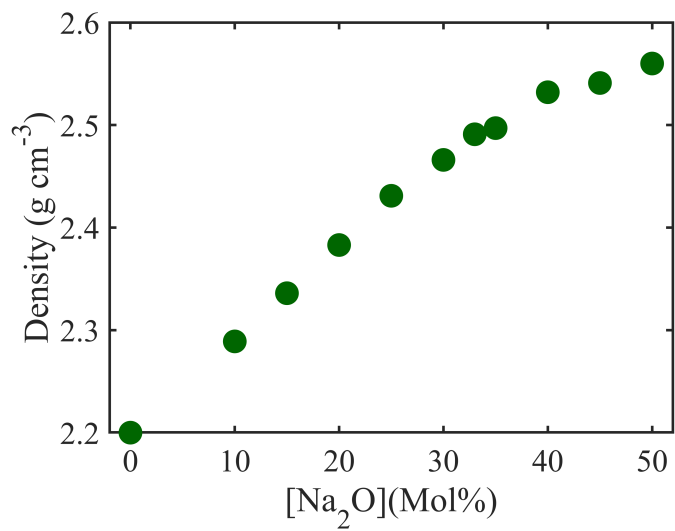

Figure S 2: Mass densities ($\rho_{\mathrm{m}}$) of $x$Na$_2$O-(100-$x$)SiO$_2$ glasses at room temperature used in simulations. Densities are adapted from Refs. [1], [2].

**Figure S3**

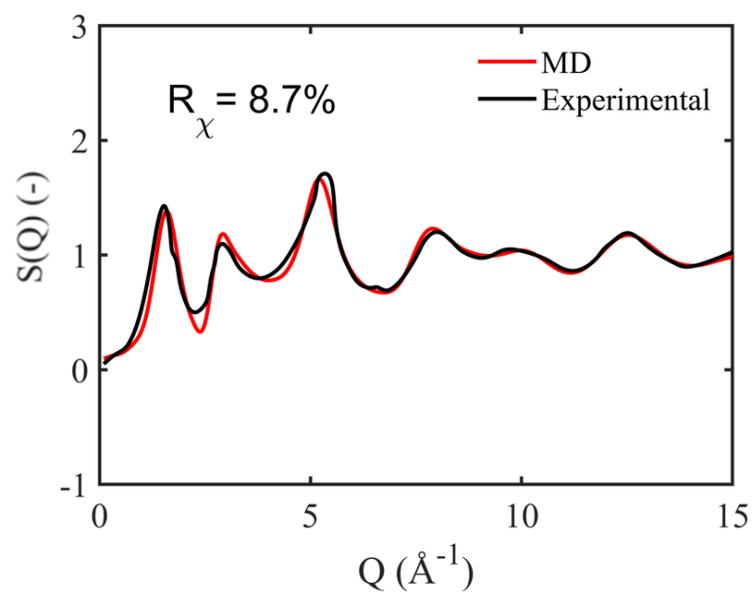

Figure S 3: Experimental structure factor from neutron diffraction experiments and simulated structure factor of $SiO_2$ glass using the potential of Teter. Based on the $R_\chi$ factor (see Methods in the main text), the medium range order shows good agreement with experimental data (generally $R_\chi < 10\%$ is considered good), yet lower agreement than that of the simulated sodium silicate glasses (Figure 1a in the main text).

**Figure S4**

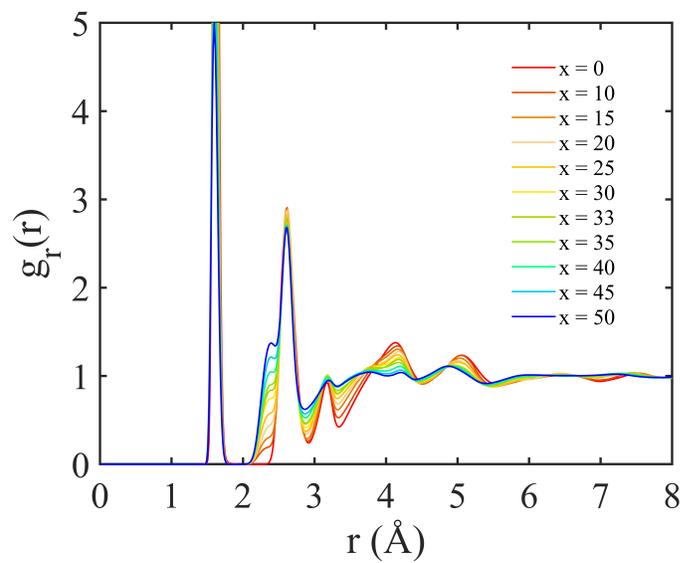

Figure S 4: Radial distribution functions (RDF's) of $x$Na$_2$O-(100-$x$)SiO$_2$ glasses with varying sodium concentrations.

**Figure S5**

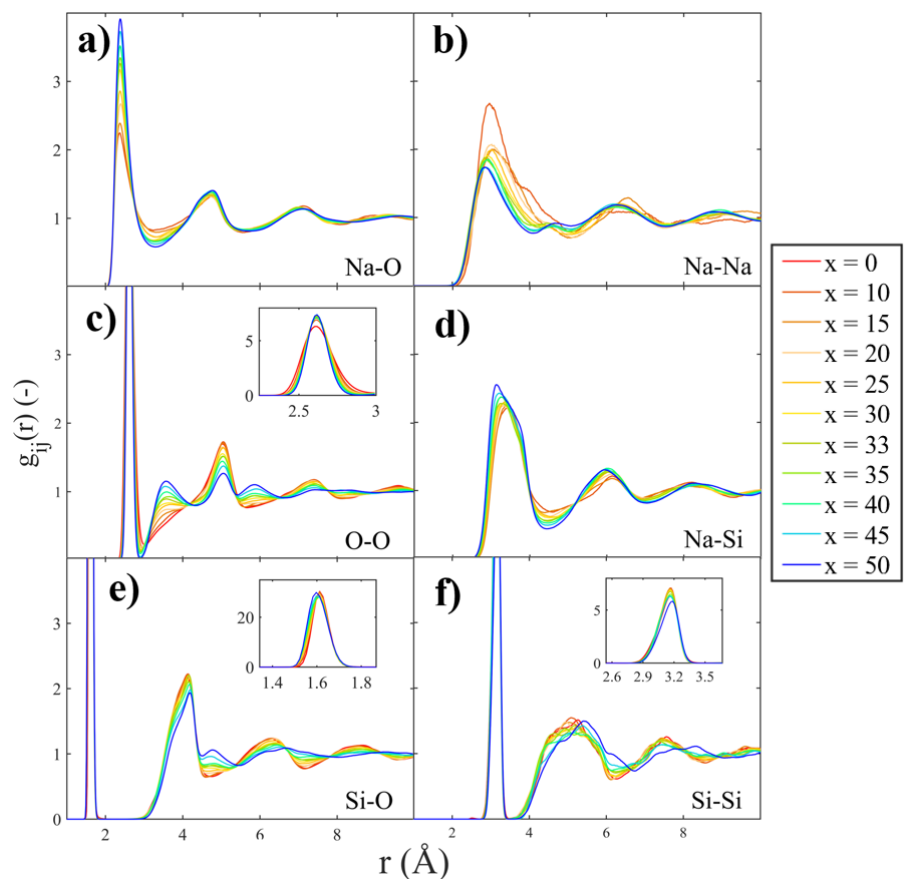

Figure S 5: Partial distribution functions (PDF's) of $x$Na$_2$O-(100-$x$)SiO$_2$ glasses with varying sodium concentrations of pair-interactions (a) Na-O, (b) Na-Na, (c) O-O, (d) Na-Si, (e) Si-O and Si-Si. Insets are provided to show peak intensities.

**Figure S6**

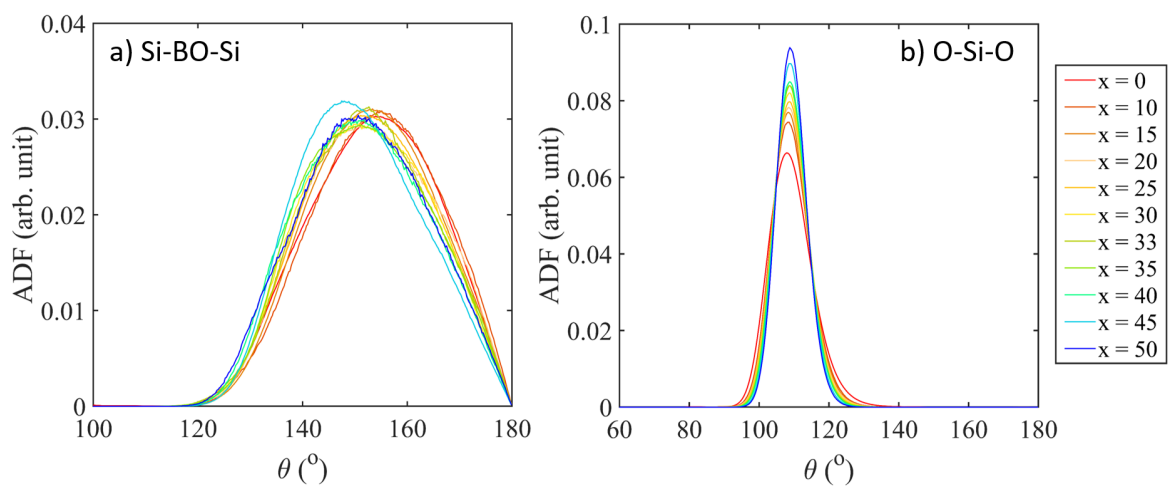

Figure S 6: Angular distribution functions (ADF's) of (a) inter-tetrahedral Silicon-Bridging oxygen-Silicon bond angles and (b) O-Si-O intra-tetrahedral bond angle distributions with increasing sodium concentration.

**Figure S7**

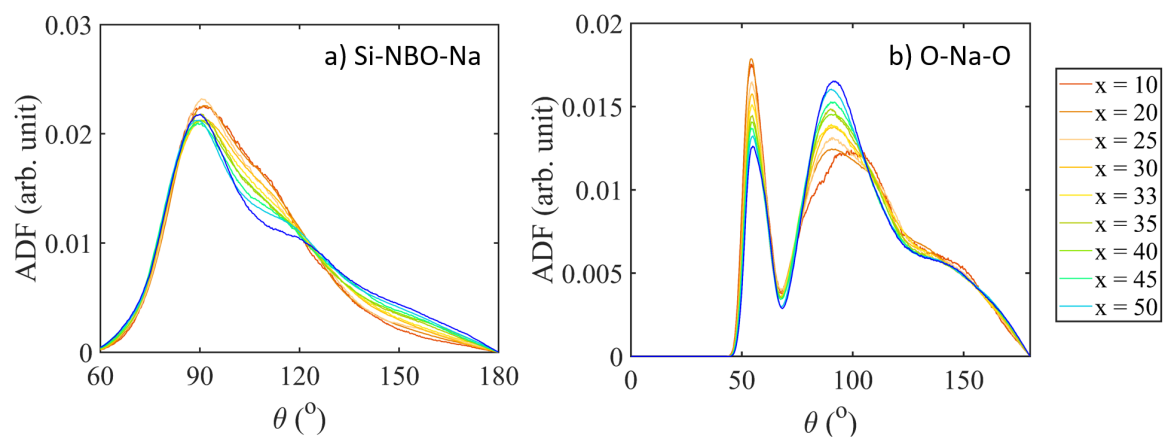

Figure S 7: Angular distribution functions (ADFs) of (a) the Si-non-bridging oxygen-Na angle formed by non-bridging oxygens (NBOs) between silicon and sodium atoms and (b) the O-Na-O angles formed in the network modifier environments.

**Figure S8**

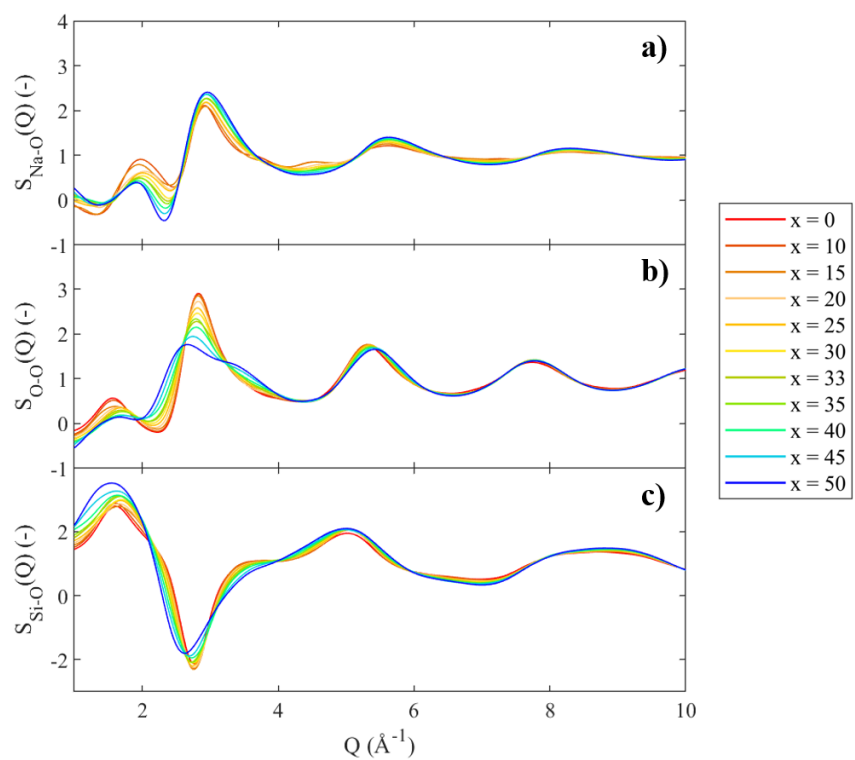

Figure S 8: Partial Faber-Ziman structure factors of pair interactions (a) Na-O, (b) O-O and (c) Si-O

**Figure S9**

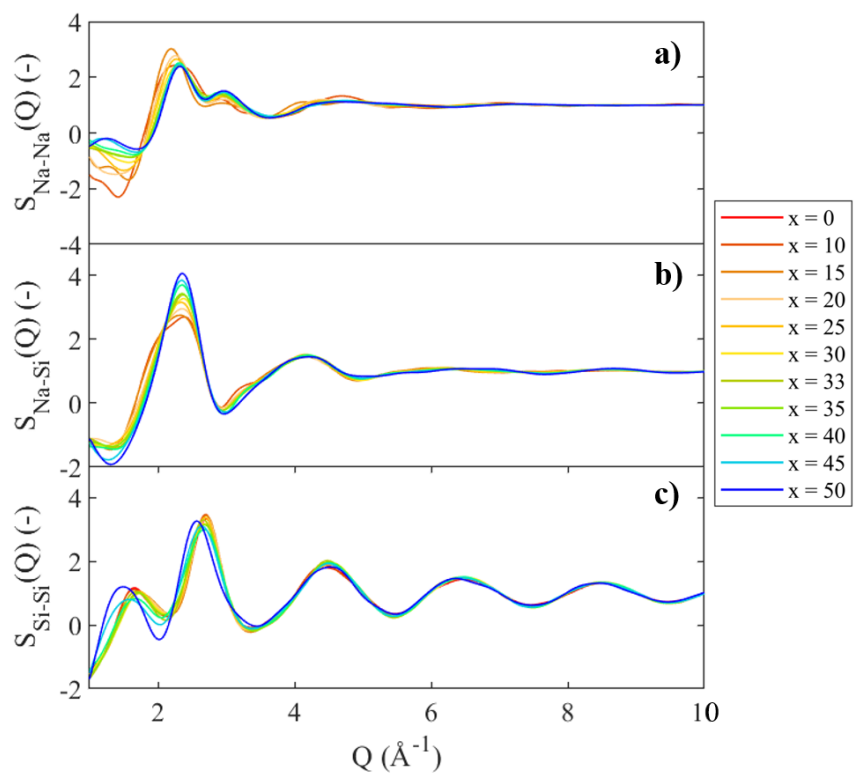

Figure S 9: Partial Faber-Ziman structure factors of cationic interactions (a) Na-Na, (b) Na-Si and (c) Si-Si.

**Figure S10**

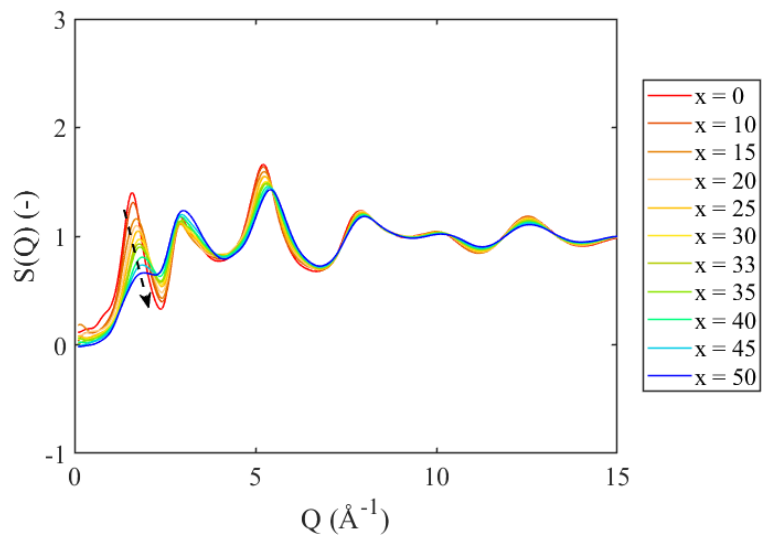

Figure S 10: Neutron-weighted structure factor of simulated glasses where $x$ is the molar concentration of $Na_2O$ ($x$) in the $xNa_2O$-$(100-x)SiO_2$ glasses. The dashed arrow is shown to emphasize the peak and positional change of the first sharp diffraction peak (FSDP), that is, the scattering peak at the lowest value of $Q$.

**Figure S11**

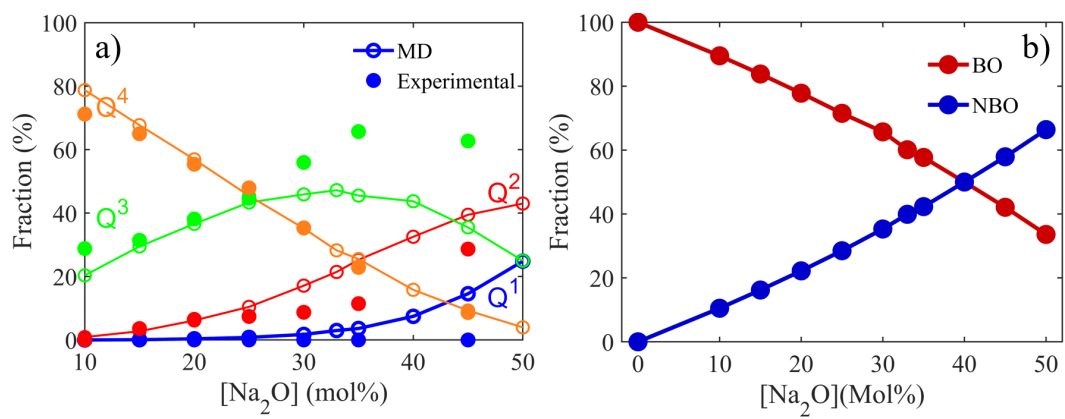

Figure S 11: Simulated and experimental data of (a) $Q^n$-distributions in glasses of 3000 atom systems and (b) the distribution of bridging and non-bridging oxygen (BO and NBO, respectively) atoms. Lines in both plots are shown as guides for the eye.

**Figure S12**

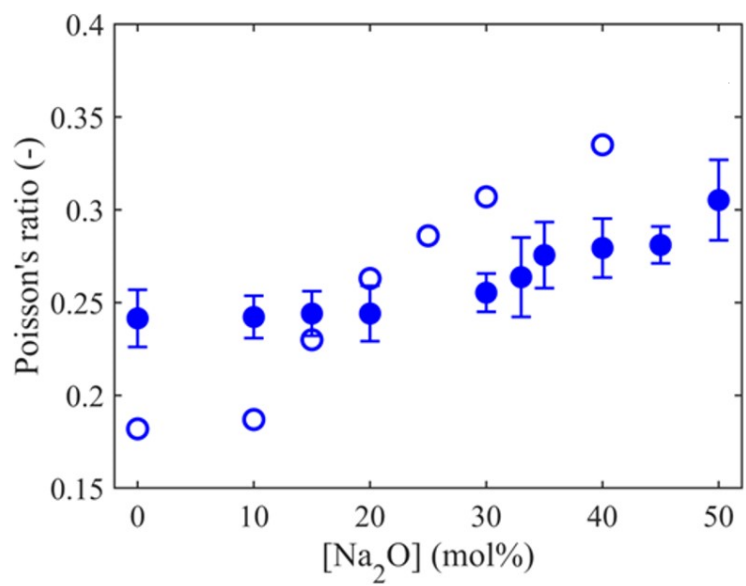

Figure S 12: Simulated and experimental Poisson's Ratio with increasing sodium content in the studied $x$Na$_2$O-(100-$x$)SiO$_2$ glasses. Simulations are somewhat able to predict the right trend, yet the increase in Poisson's ratio above 10 mol% Na$_2$O in the experimental is less pronounced in simulated data and values generally deviate more than the other moduli data.

**Figure S13**

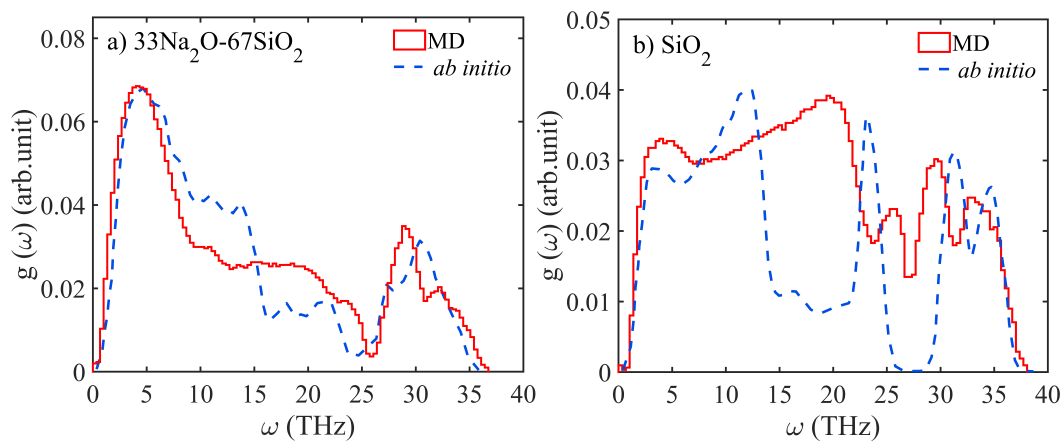

Figure S 13: Simulated vibrational density of states (VDOS) of (a) the 33.3Na$_2$O-66.7SiO$_2$ glass and (b) SiO$_2$ glass. Simulated VDOS (red) is compared with *ab initio* calculations (dashed blue) from ref. [5] showing good agreement for the sodium silicate glass, but poor agreement for SiO$_2$ glass when simulated with the potential of Teter.

**Figure S14**

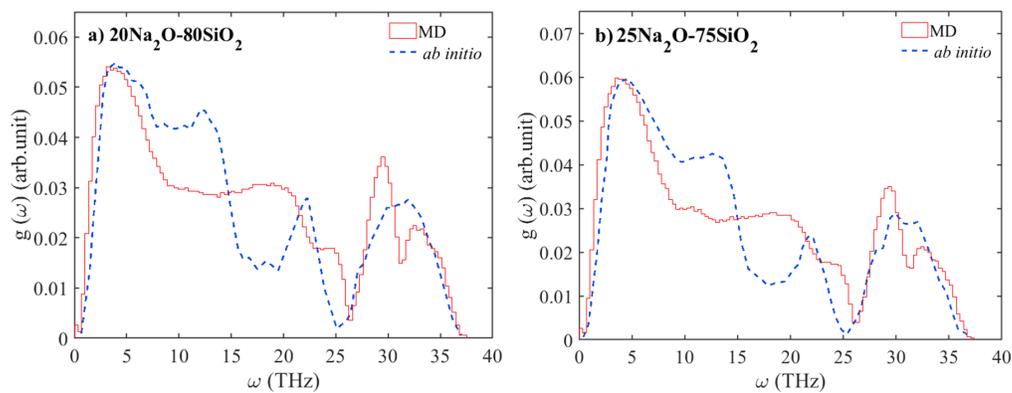

Figure S 14: Simulated vibrational density of states (VDOS) of (a) the 20Na$_2$O-80SiO$_2$ glass and (b) 25Na$_2$O-75SiO$_2$ glass. Simulated VDOS (red) is compared with *ab initio* calculations (dashed blue) from ref. [5] showing fairly high agreement for both sodium silicate glass systems.

**Figure S15**

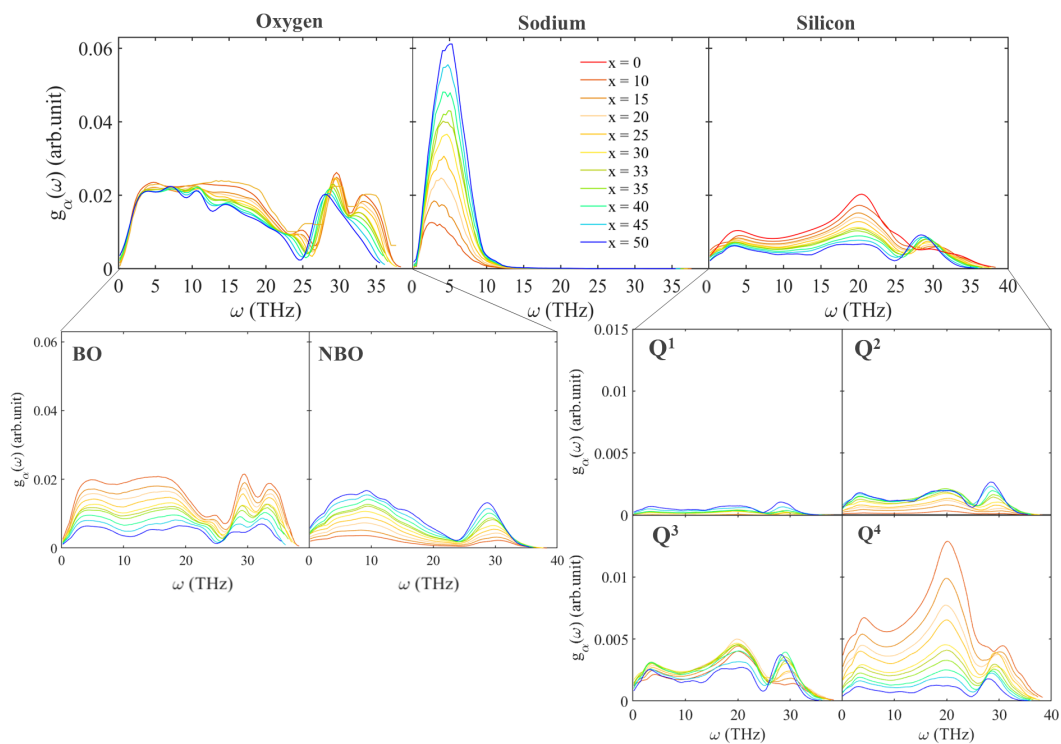

Figure S 15: Partial vibrational density of states (pVDOS) of oxygen, sodium, silicon and further subdivisions of bridging and non-bridging oxygen (BO and NBO) as well as based on $Q^n$ value, showing the frequency ranges where different atomic species contribute to the vibrational density of states.

**Figure S16**

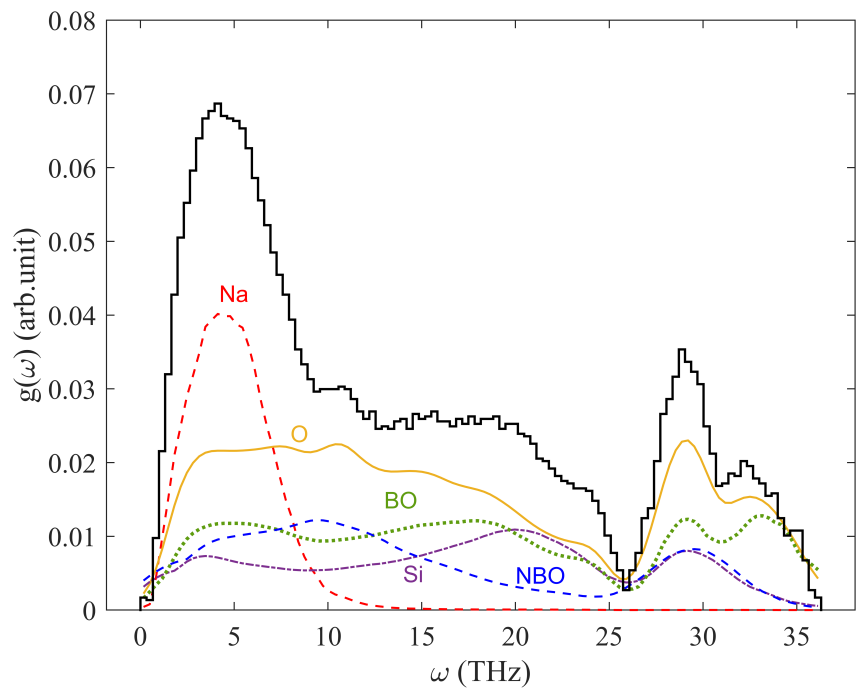

Figure S 16: (Partial) vibrational density of states (VDOS) of the 33.3Na$_2$O-66.7SiO$_2$ system illustrating the contribution to the total VDOS from atomic weighted vibrational density of states from various atomic species. We note that the contribution of O is also encompassed in the BO and NBO contributions.

**Figure S17**

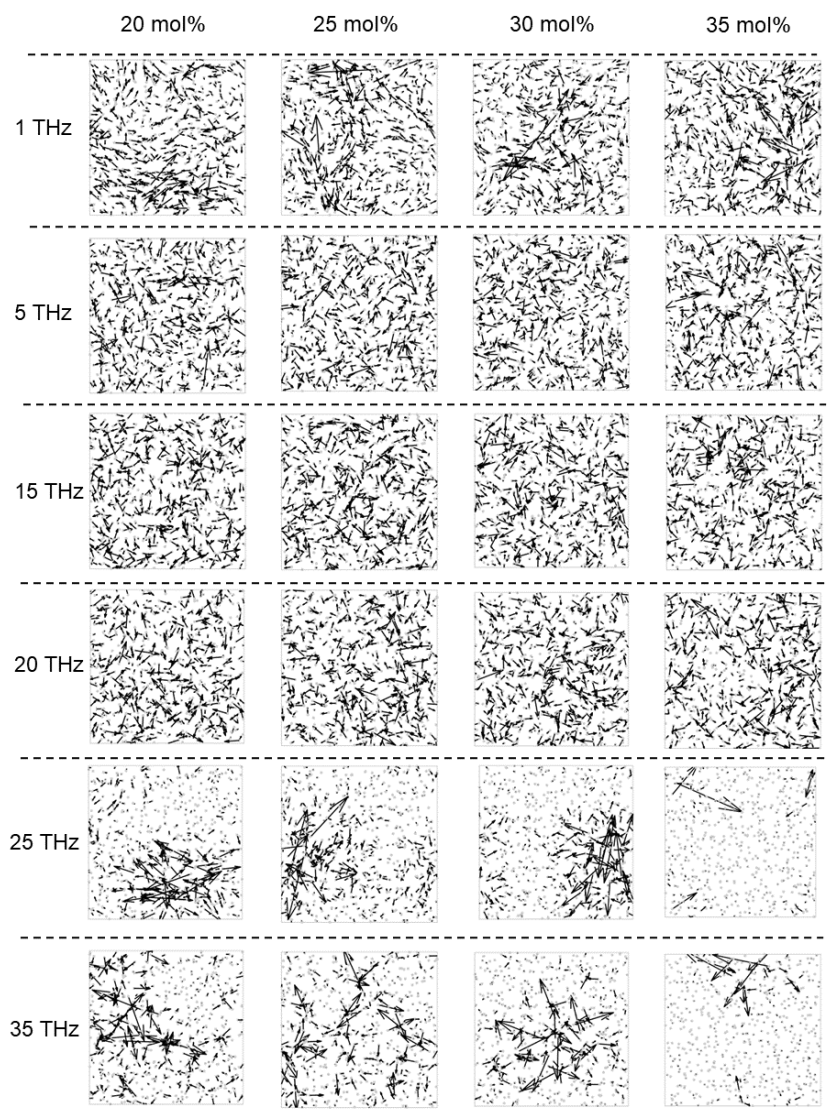

Figure S 17: Figure illustrating the changes of eigenvectors at different frequencies in glasses with increasing sodium content. Observing the low-frequency illustrations, i.e. 1 and 5 THz, eigenvectors seem to become more evenly distributed which correlates with an increasing participation ratio. The eigenvectors at 1 THz seem to become less periodic with increasing sodium contents, which could be a potential reason for the decreased per-mode heat transfer in this range, despite the increased participation ratio. In the mid-frequency range from 15-20 THz, eigenvector lengths becomes shorter in specific spatial regions of the glasses, meaning that modes become more localized with increasing sodium content. This correlates with the decreased participation ratio in the observed ranged. Lastly, high-frequency modes seem become even more localized when increasing the sodium content, suggesting that network depolymerization induces high-frequency mode localiztion.

**Figure S18**

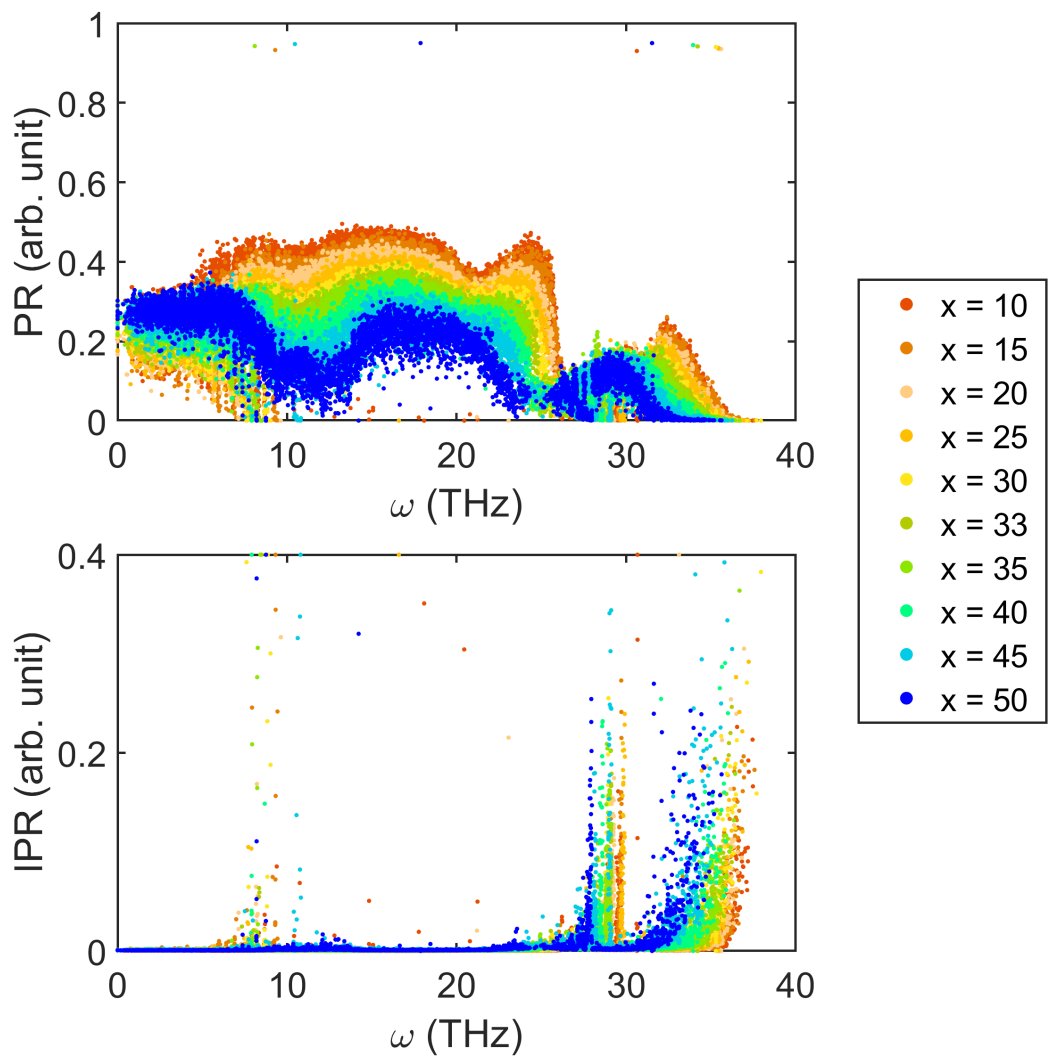

Figure S 18: Figure illustrating (a) participation ratio (PR) and (b) normalized inverse participation ratio (IPR) of simulated $x$Na$_2$O-(100-$x$)SiO$_2$ with increasing sodium content.

**Figure S19**

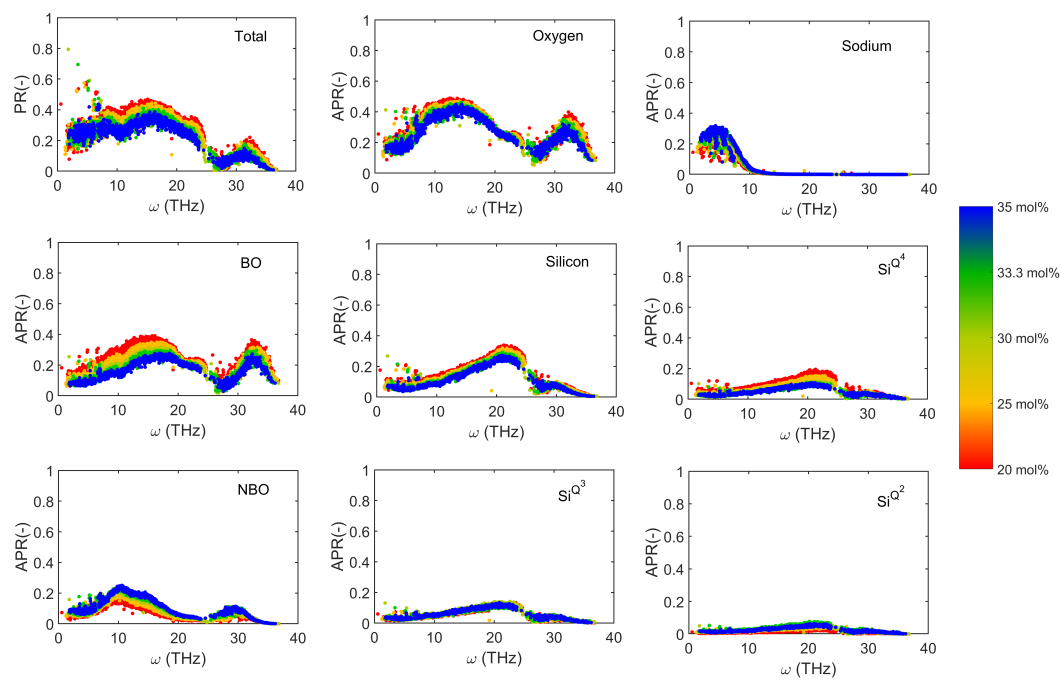

Figure S 19: Figure illustrating the total participation ratio (top left), and the atomic participation ratio (APR) of various atomic species and further subdivisions of bridging and non-bridging oxygen (BO and NBO) as well as based on $Q^n$ value of the studied $x$Na$_2$O-(100-$x$)SiO$_2$ glasses with increasing sodium content.

**Figure S20**

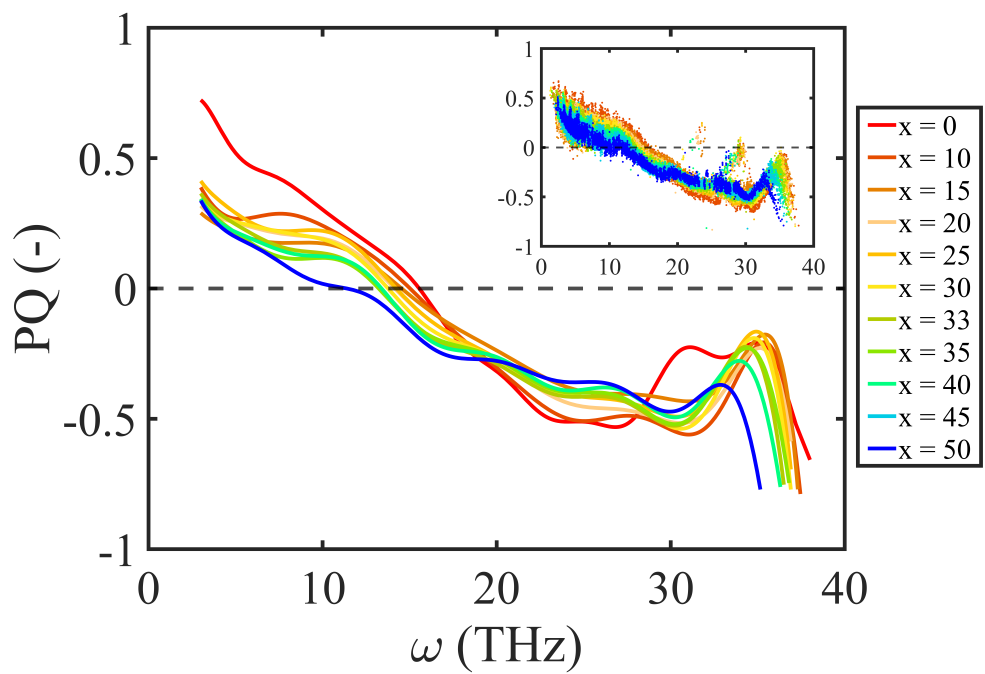

Figure S 20: Phase quotient (PQ) of simulated $x$Na$_2$O-(100-$x$)SiO$_2$ glasses with increasing sodium contents ($x$). The dashed line indicates the transition point from in-phase to out of phase vibrations ($PQ = 0$). The main plot is fitted to the data in the inset to provide easier comparison between the molar compositions of glasses.

**Figure S21**

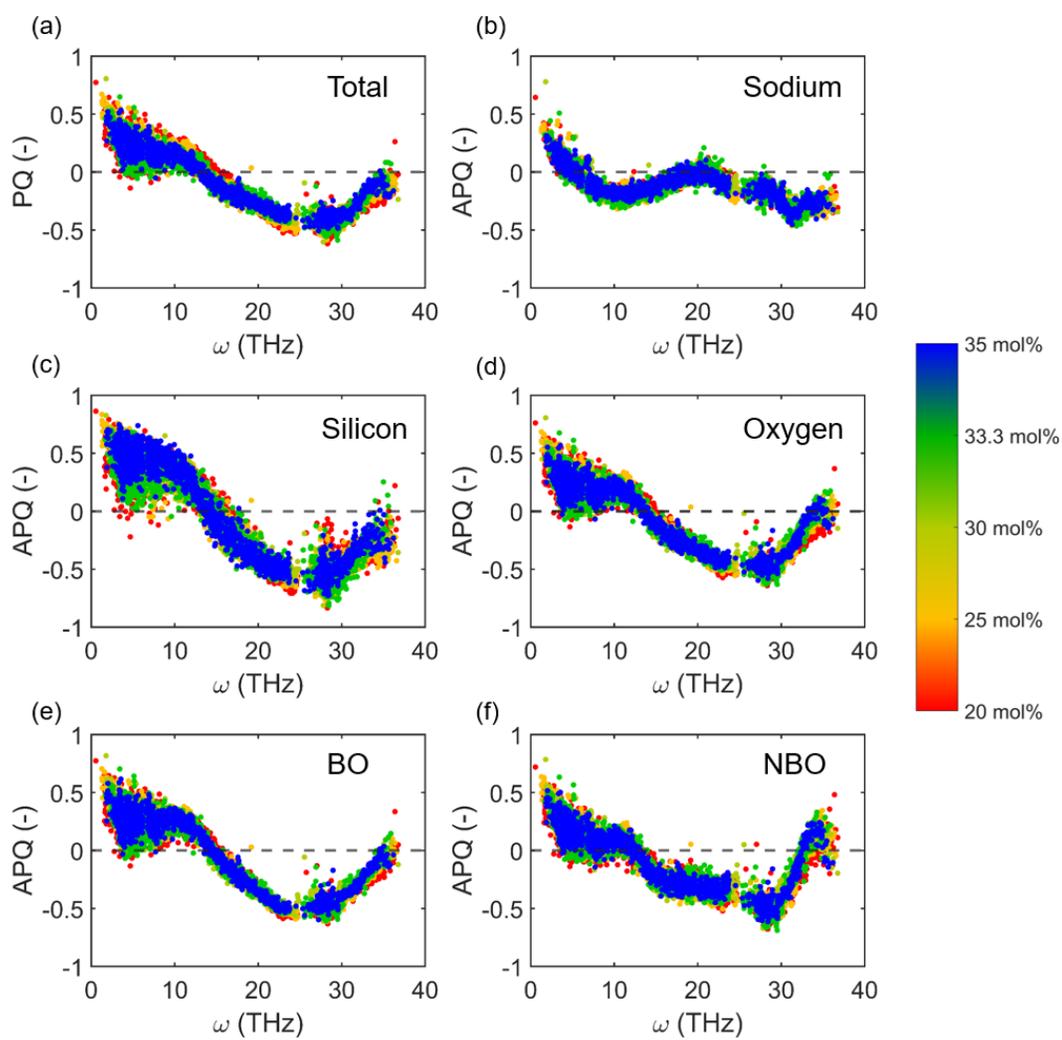

Figure S 21: Figure illustrating (a) the total phase quotient (PQ) and the atomic phase quotient (APQ) of (b) sodium, (c) silicon and (d) oxygen, (e) bridging-oxygens, and (f) non-bridging oxygens in simulated $x$Na$_2$O-(100-$x$)SiO$_2$ glasses with increasing sodium content.

**Figure S22**

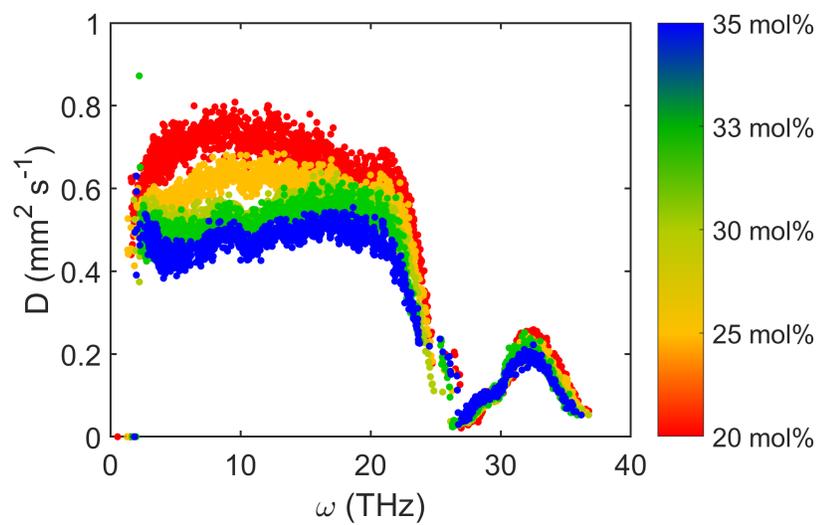

Figure S 22: Modal diffusivity, $D_i$, of simulated $x$Na$_2$O-(100-$x$)SiO$_2$ glasses with increasing sodium content.

**Figure S23**

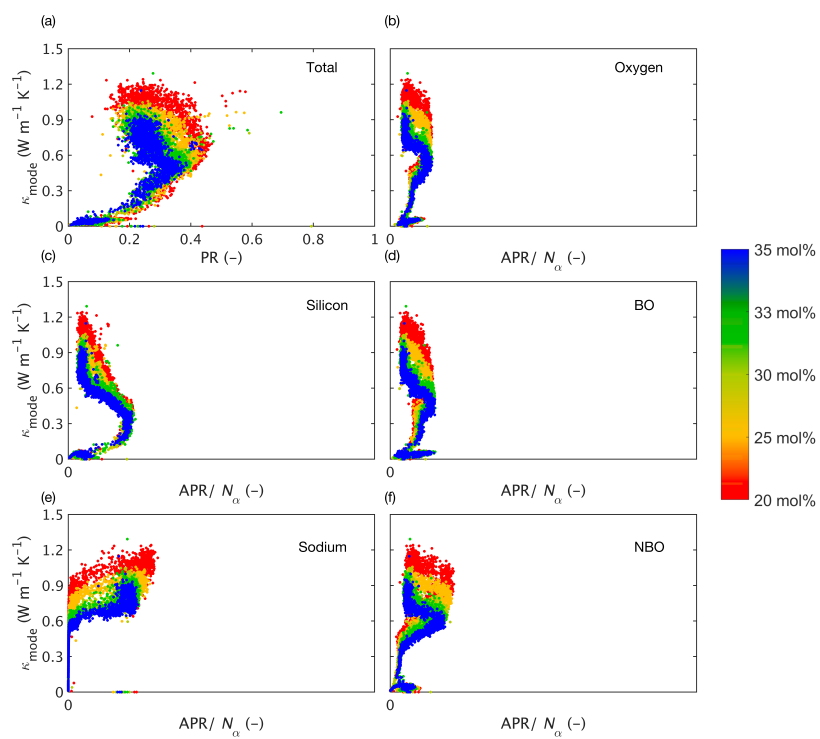

Figure S 23: Effect of sodium concentration on (a) modal participation ratio and normalized atomic participation ratio of (b) oxygen, (c) silicon, (d) bridging-oxygens, (e) sodium and (f) non-bridging oxygens coupled with modal thermal conductivities estimated with the QHGK method.



\end{document}